\documentclass[prb,twocolumn,nopreprintnumbers,amsmath,amssymb]{revtex4}
\usepackage{graphicx}
\usepackage{dcolumn}
\usepackage{bm}

\begin{document}

\title{Quantum initial condition sampling for linearized density matrix dynamics: Vibrational pure dephasing of iodine in krypton matrices}

\author{Z. Ma}
\altaffiliation{Submitted in partial fulfillment of the requirements for the Ph.D. degree in Physics at Boston University}
\author{D. F. Coker}%
\affiliation{%
Department of Chemistry, Boston University,\\
590 Commonwealth Avenue, Boston,
MA 02215
}%

\begin{abstract}
This paper reviews the linearized path integral approach for computing time dependent properties of systems that can be approximated using a mixed quantum-classical description. This approach is applied to studying vibrational pure dephasing of ground state molecular iodine in a rare gas matrix. The Feynman-Kleinert optimized harmonic approximation for the full system density operator is used to sample initial conditions for the bath degrees of freedom. This extremely efficient approach is compared with alternative initial condition sampling techniques at low temperatures where classical initial condition sampling yields dephasing rates that are nearly an order of magnitude too slow compared with quantum initial condition sampling and experimental results. 
\end{abstract}

\maketitle

\section{Introduction}
Detailed analysis of the vibrational spectroscopy of chromophores in solution can be used to investigate intermolecular interactions in condensed phases.  The theory of vibrational pure dephasing and its contribution to spectral line shapes and shifts has been worked out in detail in various limits. Perturbation theory (see references \cite{Zewail:VibDe1,Zewail:VibDe2,Zewail:Relax,Skinner02} and the and the literature cited therein), which assumes that the interaction between vibrational degrees of freedom and the environment is weak, gives an expression that enables the pure dephasing time to be computed from the zero frequency component of the time correlation function of the fluctuations in the energy gap between the vibrational levels. In many experiments, however, there may be strong initial environmental interactions associated with how the system is initially excited so techniques to study dephasing and dissipation beyond the limits of perturbation theory are important, and several non-perturbative ways to study vibrational dynamics have appeared in the literature \cite{Skinner01,Skinner02,Skinner04,Weiss,Makarov99}. One very fruitful way to go beyond perturbation theory, for example, is to employ an idealized model which can be worked out analytically  such as a two level system appropriately coupled to a harmonic bath for which the effects of environmental dephasing on lineshape and spectral shifts have been worked out in detail \cite{Skinner01,Skinner02,Skinner04}. Several general predictions emerge from this analysis concerning, for example, how lineshapes are affected by characteristics of bath spectral density, and temperature, etc. Experimentalists can use the predictions of this theory to interpret their findings in terms of the nature of the underlying interactions in condensed phase systems. The interactions between the vibrational coordinate and the environment can depend on vibrational excitation in a complex way and this can complicate use of the two level system theory described above. Also employing a harmonic bath when anharmonicity may be significant could make the use of such a theory questionable.

An alternative non-perturbative method that has received considerable attention in the quantum optics literature \cite{Dalibard92,Dum92} and recently has become a focus in molecular science applications \cite{Makarov99} is the so-called stochastic wave function approach \cite{Weiss}. With this method the evolution of the density matrix is assumed to take a simplified Bloch form parameterized by phenomenological coefficients governing the decay of coherence and the transfer of population. Rather than propagating the $n \times n$ density matrix elements, a stochastic wave function represented in terms of $n$ relevant basis functions is evolved using statistical rules designed in such a way so that the reduced density constructed from an ensemble of evolved stochastic wave functions reproduces the evolution of the Bloch equations. Since this approach relies on a Bloch model form containing phenomenological parameters it thus provides an efficient, linear scaling approach for fitting experimental data to such models. As these models are often discussed in terms of gas phase collision processes, interpreting such fits in a meaningful way and extracting information about a microscopic mechanism of the relevant decoherence processes operating in condensed phase systems is not straightforward. 

A further possible alternative approach is the use of a realistic microscopic model with vibrational state dependent interactions as has been developed in various contexts over the last few years \cite{Gerber96,Gerber97,Gerber99}. Such an approach requires a quantum dynamical treatment of dephasing and microscopic simulation methods to address this problem have recently been developed \cite{Poulsen03,ShiGeva03b,ShiGeva03c,ShiGeva03d,Hernandez98,Miller01-IVR,Coker05b,Martens01}. In this paper we extend these methods to include quantum initial condition sampling that should accurately capture the relevant underlying spectral density of the realistic model system. Thus unlike the theories discussed above which are built on use of different models (e.g. Debye or pseudolocal mode spectral densities), with the approach employed here we do not need to assume anything about the spectral density arising from the interactions. As such direct, model free, interpretation of the experimental results is possible. 

The specific set of experiments which we will study with this alternative approach come from recent work from Apkarian and coworkers \cite{Apkarian00,Apkarian01a,Apkarian01b,Apkarian04a,Apkarian04b,Apkarian04c,Apkarian05a,Apkarian05b,Apkarian05c} in which they use Time Resolved Coherent Anti-Stokes Raman Scattering (TR-CARS) to directly probe the dephasing of vibrationally excited I$_2$ wave-packet components due to interactions of this chromophore with a rare gas matrix. This well studied system is chosen as a benchmark on which to evaluate the approach due to the availability of accurate interaction potential data. Also, as distinct from traditional frequency domain experiments in which lineshapes and frequency shifts are indirectly interpreted in terms of the evolution of the density matrix, these time domain experiments can be directly connected to the theory of decoherence which we outline below.  

The phenomenon of vibrational dephasing can be probed when a vibrational subsystem, described by Hamiltonian $\hat{H}_s$, with eigenstates given by $\hat{H}_s |v_0\rangle = \epsilon_{v_0} |v_0\rangle$, is prepared in some coherent superposition state, $|\psi(0)\rangle = \sum_{v_0} c_{v_0} |v_0\rangle$, for example, in the presence of an environmental subsystem in state $|\chi(0)\rangle$ which is unaffected by this preparation. Thus the composite system initial state can be described by a separable product state $|\Psi (0)\rangle = |\psi(0)\rangle |\chi(0)\rangle$. Important deviations from this separable product initial condition in the limit of strong system - environment coupling have been discussed \cite{Pechukas94,Laird91a,Laird91b}. If the quantum subsystem and the environment are uncoupled, the initial coherent superposition of vibrational states will be maintained as the whole system evolves, {\em i.e.} the amplitudes and relative phases of the component state contributions will be constant, and the composite system wave function will remain separable. In the more general situation, however, the dynamics of the full system is governed by the coupled system-bath Hamiltonian, $\hat{H} = \hat{H}_s + \hat{H}_b + \hat{H}_{s-b}$ and the above initially separable wave function will evolve into an entangled state $|\Psi(t) \rangle = \sum_{v_0} c_{v_0} \exp[-{i\over \hbar}\hat{H} t] |\chi(0) v_0 \rangle = \sum_{v_N} |\chi_{v_N}(t) \rangle |v_N\rangle$ in which the amplitudes and relative phases of the different vibrational state contributions will change with time.  In the last equality in the previous result we have projected the composite state onto the diabatic vibrational basis states $|v_N\rangle$, and the bath states that appear in this entanglement, labelled according to this basis, are obtained as $| \chi_{v_N}(t) \rangle = \langle v_N | \sum_{v_0} c_{v_0} \exp[-{i \over \hbar} \hat{H}t] |\chi(0) v_0 \rangle$. The timescale for the variation in the phase of the different vibrational state components is known synonymously as the vibrational dephasing or decoherence time. The variation in the amplitude, on the other hand, reflects the vibrational state population relaxation time. With this description, the bath states $|\chi_{v_N}(t) \rangle$ contain all the relative phase and amplitude information of the contributions from the different vibrational basis states. With the view outlined here we are assuming that the details of the shapes of the exciting radiation field pulses can be neglected and that we can focus our attention on the evolution of the reduced density matrix. A complete description of the experiments would require considering details of the time dependent interactions of the full system with the radiation field and is beyond the scope of the current investigation of vibrational pure dephasing times. 

In the vibrational pure dephasing limit it is assumed that the composite system dynamics results in slow amplitude relaxation of the chosen vibrational basis states on the fast timescale of the fluctuations in the phases of the different component basis states. Thus we suppose that contributions to final state $v_N$ amplitude that originate from initial states $v_0$, different from $v_N$, are negligible and we approximate the above expression as $|\chi_{v_N}(t) \rangle \approx c_{v_N} \langle v_N | \exp[-{i \over \hbar} \hat{H} t | \chi(0) v_N \rangle$. This is equivalent to the vibrationally adiabatic approximation. The timescale for the vibrational pure dephasing process is governed by the distribution of fluctuations in the phase of the state $|\chi_{v_N}(t) \rangle$ which is determined by two factors: (1) the strength of the interactions between the vibrational subsystem and its environment which will in general depend on the particular vibrational state $|v_N\rangle$, and (2) the distribution of initial states of the environment $|\chi(0) \rangle$. This paper develops a general approach for computing vibrational pure dephasing rates in condensed phase systems which incorporates both the varying strength of environmental interactions with quantum subsystem state, and the effect of quantum dispersion on the nature of the distribution of the initial environmental states. 

The quantum vibrational state dependent intermolecular potential idea developed in the late 1990's \cite{Gerber96,Gerber97,Gerber99} provides a framework for the representation that we employ here to account for the the variation of environmental interaction strength with vibrational state. Specifically the model of Martens and co-workers \cite{Martens04,Martens05a,Martens05b,Martens06} is adopted for this purpose. The main goal of the work here is to explore how thermal and quantum fluctuations of the initial environment influence vibrational pure dephasing in a model system with realistic interactions. Thus we will compare classical and approximate quantum descriptions of the thermal environmental initial distribution and explore their effects on dephasing dynamics. In addition a complete and general derivation of the Feynman-Kleinert-Wigner (FK-Wigner) approach for sampling quantum initial conditions is presented. 

The paper is organized as follows: First we outline a general approximate approach for treating the evolution of the vibrational reduced density matrix. With this linearized approximation, quasi-classical trajectories are evolved from quantum initial conditions sampled from the Wigner transform of the initial density. The main methods section of the paper summarizes the FK-Wigner approach used for this sampling. Details of the complete derivation of this approach are given in the appendices. The final methods subsection makes comparisons of this approach with other published techniques. Finally in section \ref{sec:Results} we outline the implementation of these methods for application to computing vibrational pure dephasing rates and compare these data computed using different approximations with available experimental results. Concluding remarks are given in section \ref{sec:conclusion}.

\section{Methods}

\subsection{Vibrational Pure dephasing}
The density matrix formulation offers a convenient way to rewrite the wave function description outlined above in a useful form for making further approximations and interpretations. As noted earlier, the experiments of interest involve exciting the vibrational subsystem independently of the environment. The bath degrees of freedom are thus assumed to be initially prepared in thermal equilibrium with the quantum subsystem. The quantum vibrational subsystem is then prepared by rapid pulsed laser excitation, for example, in a non-equilibrium coherent superposition of excited vibrational states $|\psi(0) \rangle$ as described in the Introduction. We assume that the non-equilibrium composite system is thus initially described by a density operator which is a product form $\hat{\rho}(0)=\hat{\rho}_s(0) \hat{\rho}_{b}^{e}(0)$, where $\hat{\rho}_{b}^{e}$ is the bath part of the equilibrium density operator for which we will develop approximations, and the non-equilibrium quantum subsystem density for the initially excited coherent superposition state is $\hat{\rho}_s(0) = |\psi(0) \rangle \langle \psi(0)|$, which has various component operators, for example, $c_{v_0} c_{v_0'}^* |v_0\rangle \langle v_0'|$, that depend on what states of the quantum subsystem are coherently excited by the laser pulses. 

The coherently excited composite system will evolve from this factored initial state to an entangled state as a result of the coupled full system evolution as discussed above. This entanglement will thus be described by the full system time dependent density matrix with elements in the environmental coordinate and vibrational subsystem state representation given as $\langle R_N, v_N | \hat{\rho}(t) | R_N', v_N' \rangle$ $= \langle R_N, v_N | e^{-i \hat{H} t / \hbar} \hat{\rho}(0) e^{i \hat{H} t / \hbar} | R_N', v_N' \rangle$. The experiments of interest probe only the quantum subsystem states so we will study the reduced density matrix elements obtained by tracing over all the bath degrees of freedom {\em i.e.} $\rho_{v_N v_N'}^{red}(t)= \int dR_N \langle R_N, v_N |\hat{\rho}(t)| R_N, v_N' \rangle$. Suppose the preparation selects out the $\hat{\rho}_s(0) = |v_0 \rangle \langle v_0'|$ component sub-system density operator initially, then the reduced density operator matrix elements at time $t$ will have the form 
\begin{eqnarray} 
\rho_{v_N,v_N'}^{red}(t)  & = & \int dR_N \int dR_0 \int d R_0'  \langle R_N, v_N | e^{-i \hat{H} t /\hbar} | R_0, v_0 \rangle \nonumber\\ 
\ & \ & \times \langle  R_0 | \hat{\rho}_b^e | R_0' \rangle \langle R_0', v_0' |  e^{i \hat{H} t /\hbar} |R_N, v_N' \rangle \label{eq:red-den}
\end{eqnarray}
and involve forward and backward propagator matrix elements as well as initial bath density operator matrix elements. 

Suppose the full Hamiltonian is
\begin{equation}
\hat{H} = {\hat{p}_s^2 \over 2m} + v(\hat{s}) + {\hat{P}^2 \over 2M_b} + V_b(\hat{R}) + \Phi_{s-b}(\hat{s},\hat{R}) 
\end{equation} 
which can be expressed in our quantum subsystem diabatic vibrational basis as $\hat{H} = \hat{P}^2/ 2M_b + \sum_{\alpha,\beta} |\alpha \rangle h_{\alpha \beta}(\hat{R}) \langle \beta|$ where $h_{\alpha \beta}(\hat{R}) = [\epsilon_{\alpha} + V_b(\hat{R})] \delta_{\alpha \beta} + \langle \alpha | \Phi_{s-b} | \beta \rangle (\hat{R})$. Here $\Phi_{s-b}$ is the system-bath interaction potential. In the case of vibrational pure dephasing we suppose that $h_{\alpha \beta}(\hat{R}) \sim 0$ for $\alpha \ne \beta$, {\em i.e.} the off-diagonal elements are small compared to the diagonal elements, so we can approximate the full system Hamiltonian by the diagonal form $\hat{H}_d =  \hat{P}^2 / 2M_b + \sum_{\alpha} |\alpha \rangle h_{\alpha \alpha}(\hat{R}) \langle \alpha|$ for all important $R$. In this case there is no population relaxation between our vibrational basis states so the propagator matrix elements appearing in Eq.(\ref{eq:red-den}) can be written in the composite bath subsystem path integral forms as follows:
\begin{eqnarray}
&&\langle R_N, v_N | e^{-i \hat{H} t /\hbar} | R_0, v_0 \rangle\nonumber\\
&& =  \delta_{v_N,v_0} \int_{R(0)=R_0}^{R(t)=R_N} {\cal D}[R(t)] e^{{i \over \hbar} S_{v_0 v_0}[R(t)]} 
\end{eqnarray}
and 
\begin{eqnarray}
&&\langle R_0', v_0' | e^{i \hat{H} t /\hbar} | R_N, v_N' \rangle \nonumber \\
&&=  \delta_{v_N',v_0'} \int_{R'(0)= R'_0}^{R'(t)=R_N} {\cal D}[R'(t)] e^{-{i \over \hbar} S_{v'_0 v'_0}[R'(t)]} 
\end{eqnarray}
where the forward path action, for example, is 
\begin{eqnarray}
&&S_{v_0, v_0}[R(t)] = \int_0^t dt' \{ {1 \over 2}M_b \dot{R}^2(t') \nonumber\\ 
&&- [\epsilon_{v_0} + V_b(R(t')) + \langle v_0 | \Phi_{s-b} | v_0 \rangle(R(t'))] \}
\end{eqnarray}
and a similar expression for the action along the backward path is obtained by modifying the vibrational state accordingly to $v'_0$.

Combining these expressions, the reduced density matrix in the pure dephasing limit can be written as
\begin{widetext}
\begin{eqnarray}
\rho_{v_0, v'_0}(t) & = & \int dR_N \int dR_0 \int dR'_0 \langle R_0 |\hat{\rho}^e_b |R'_0 \rangle \label{eq:redrho}
  \int_{R(0)=R_0}^{R(t)=R_N} {\cal D}[R(t)] \int_{R'(0)=R'_0}^{R'(t)=R_N} {\cal D}[R'(t)]
  e^{{i \over \hbar}  \{S_{v_0, v_0}[R(t)]-S_{v'_0, v'_0}[R'(t)] \} }
\end{eqnarray} 
\end{widetext}
In the above expression the bath evolution is still described at the full quantum level. To proceed to a computable expression for the time dependence of the reduced density matrix elements we follow  various authors \cite{Poulsen03,ShiGeva03b,ShiGeva03c,ShiGeva03d,Hernandez98,Miller01-IVR} 
and combine the forward and backward propagators in Eq.(\ref{eq:redrho}) and write the product in terms of mean, $\bar{R}(t) = [R(t)+ R'(t)]/2$, and difference, $Z(t) = R(t)- R'(t)$, bath subsystem path coordinates (with a similar transformation for the bath momenta, $\bar{P}(t)=[P(t)+ P'(t)]/2$ and $Y(t) = P(t)-P'(t)$). Next, the phase of the integrand in Eq.(\ref{eq:redrho}) is expanded in the path difference variables. In condense phase systems various arguments can be given to justify keeping only low order terms in the path difference \cite{CaldeiraLeggett:QBM, CaldeiraLeggett:QTinDS,CaldeiraLeggett:Exact,Poulsen03,Coker05b}. 
Thus we proceed by truncating the expansion of the phase to linear order in the path difference variables. With this approximation the integrals over the initial difference coordinate $Z_0$ can be performed defining the Wigner transform of the initial density operator 
\begin{equation}
( \hat{\rho}_b^e)_W ({\bar R}_0,{\bar P}_1)=\int d Z_0 \langle {\bar R}_0 + {Z_0 \over 2} | {\hat \rho}_b^e | {\bar R}_0 - {Z_0 \over 2} \rangle e^{-{i} {\bar P}_1 Z_0}
\end{equation}
If we discretize the paths appearing in Eq.(\ref{eq:redrho}) by inserting $N$ resolutions of the identity in the bath subsystem phase space and take small time steps $\delta t = t/N$, thus writing the discrete path variables as $R_k = R(k\delta t)$, for example, we find that all integrals over the difference coordinates, $Z_k$, and difference momenta, $Y_k$ for $0 < k < N$ that appear in the discretized result can also be performed in the linearized approximation since they are integral representations of $\delta$-functions \cite{Coker05a,Coker05b}. Thus, the linearized approximation for the evolution of the reduced density operator becomes
\begin{widetext}
\begin{eqnarray}\label{eq:Lin-den}
\rho_{v_0,v_0'}(t) & =& \int d{\bar R}_0 \int \prod_{k=1}^{N} d{\bar R}_k{d{\bar P}_k\over 2\pi\hbar}({\hat \rho}_b^e)_W ({\bar R}_0,{\bar P}_1) e^{-{i\over \hbar}(\epsilon_{v_0}-\epsilon_{v_0'})t} e^{-{i \over \hbar} \delta t\sum_{k=1}^N \Delta \Phi_{v_0,v_0'}({\bar R}_k)}  \nonumber \\
 && \times \prod_{k=1}^{N-1} \delta \left ( {{\bar P}_{k+1}-{\bar P}_k\over \delta t}- F^{v_0,v_0'}_k \right ) \prod_{k=1}^N \delta \left ({{\bar P}_k\over M } -{{\bar R}_k - {\bar R}_{k-1}\over \delta t} \right )
\end{eqnarray}
\end{widetext}
where 
\begin{equation}\label{eq:AdiabaticForce}
F^{v_0,v_0'}_k=-{1\over 2} \left \{ \nabla_{\bar R_k} h_{v_0,v_0}({\bar R}_k)+\nabla_{\bar R_k} h_{v_0',v_0'}({\bar R}_k) \right \} 
\end{equation}
and 
\begin{equation}
\Delta \Phi_{v_0,v_0'} = \langle v_0| \Phi_{s-b} | v_0 \rangle - \langle v_0'| \Phi_{s-b} | v_0' \rangle
\end{equation}

This result indicates that the time evolution of the density operator matrix elements can be approximated by first sampling the initial environment phase space from the Wigner transform of the bath part of the thermal equilibrium density of the full system. In our low temperature calculations we approximate this equilibrium density by assuming that the environment experiences interactions with only the ground state of the vibrational subsystem. At higher temperatures excited vibrational states should be included in this initial sampling. Next, the product of $\delta$-functions in Eq.(\ref{eq:Lin-den}) indicates that with in the linearized approximation the time dependence of the density matrix elements can be obtained by evolving classical trajectories with the sampled initial conditions and subject to the mean of the forces arising from the quantum states involved in the prepared superposition state. Finally Eq.(\ref{eq:Lin-den}) gives that the contributions from each sampled trajectory to the density are obtained by adding coherently phase factors computed along these trajectories. An approach based on the same approximations outlined above has been used in other work \cite{Coker05a,Coker05b} to compute various quantum time correlation functions exploring electron transport, and vibrational energy relaxation, for example. In fact, approaches like this which employ the Wigner transform \cite{Hillery84} of the equilibrium density as the distribution of initial conditions and evolve the classical dynamics with a mean Hamiltonian have a long history in computing spectroscopic correlation functions, for example, \cite{Mukamel82,Shemetulskis92}. The work of Egorov {\em et al.} \cite{Egorov98} provides an important comparison of this type of approach with alternative classical and mixed quantum-classical methods in the context of computing model condensed phase vibronic spectra.   

\begin{figure}
\centerline{\includegraphics[width=4.in]{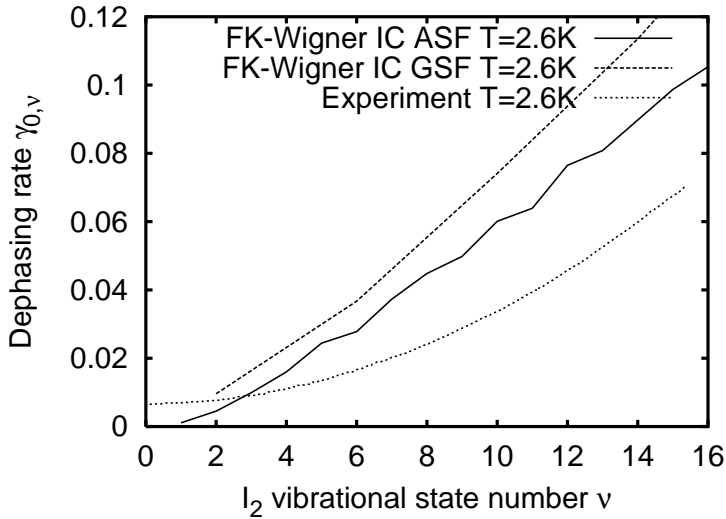}}
\caption{\label{fig:ave-gs-gamma-compare}Comparison of experimental and computed dephasing rates versus vibrational state for I$2$ in solid Krypton at low temperatures. Solid curve gives results obtained using trajectories propagated with the Average vibrational State Force (ASF), dashed curve gives results obtained with vibrational Ground State Force (GSF) (see text for discussion of these different approaches), and dotted curve displays the experimental results.}
\end{figure}

The rest of the paper describes the computational approach we adopt for sampling the Wigner density and the mean Hamiltionian dynamics calculations that we have conducted to explore the characteristics of the vibrational pure dephasing process in low temperatures crystals. In Fig. \ref{fig:ave-gs-gamma-compare} we present our dephasing results for I$_2$ in solid krypton and compare with experiments in order to demonstrate the sensitivity of these type of calculations to the potential field used in the underlying dynamics. The calculation results presented in this figure both use the FK-Wigner initial condition sampling approach detailed in the next section. The different sets of calculated dephasing rates presented here, however, were obtained using different ways of propagating the classical trajectories. The solid curve presents results obtained using the average state force (ASF) to drive the environmental dynamics as suggested by Eq.(\ref{eq:Lin-den}) and discussed above. An alternative approach is to evolve the environmental dynamics using forces arising from the ground vibrational state potential only\cite{Mukamel82,Egorov98}. The curve labeled GSF (Ground vibrational State Force) in Fig. \ref{fig:ave-gs-gamma-compare} is obtained in this way. We see that both sets of calculated dephasing rates are generally faster than the experimentally observed results. These discrepancies with experiment probably reflect inaccuracy in the model interactions employed in these calculations which we outline later. However, we generally find that the dephasing rates computed using the average surface force are in better agreement with experiment than results computed using the ground vibrational state force. The two calculated dephasing rate curves should coalesce at low vibrational state number where the average potential from states 0 and $\nu$ is essentially that of the ground state. As we increase $\nu$ the average force experienced by the environment $F^{0,\nu} = [\nabla_{\bar{R}} h_{0,0}(\bar{R}) + \nabla_{\bar{R}} h_{\nu,\nu}(\bar{R})]/2$ in our ASF calculations becomes more and more different from the ground state force $F^{0,0} = \nabla_{\bar{R}} h_{0,0}(\bar{R})$ employed in the GSF calculations and the deviation between the two curves is apparent. As discussed above, according to Eq. (\ref{eq:Lin-den}), the dephasing rates are obtained by averaging the phase factor in the energy gap between the two relevant vibrational states. If the forces governing the fluctuations in the energy gap arise from both of the states we expect that each state energy will fluctuate in a similar way leading to slower dephasing. If, on the other hand, the solvent responds to only the ground state force, as in the GSF calculations, the excited state energy may fluctuated more strongly as it does not play a role in influencing the environmental dynamics. This explains why the dephasing rates computed with the average force dynamics are slower than those obtained with the ground state force.   

The main goal of our work is to study the effect of the distribution of environmental states on vibrational pure dephasing, making particular contact with low temperature experiments. Under these conditions quantum dispersion and tunneling of the environmental degrees of freedom may play a significant role. Thus we will employ various approximations to the Wigner transform of the initial density and compute their effect of the linearized approximation to the dephasing dynamics given in Eq.(\ref{eq:Lin-den}). General methods for computing the Wigner transform of the Boltzmann operator are as yet unavailable. In our work we thus employ an approximation to this operator. First we assume that the temperature is sufficiently high, and that particle masses are large, so that identical particle statistics complications can be ignored. Next, as detailed below, we employ an approach pioneered by Feynman and Kleinert that approximates the many body Boltzmann operator assuming a locally quadratic form for the interactions. This approximate approach has recently been implemented in various condensed phase applications studying quantum effects on transport and spectroscopy by Poulsen and co-workers \cite{Poulsen03, Poulsen04, Poulsen05}. Our study will compare vibrational dephasing results obtained using this approximate Boltzmann operator which can incorporate some quantum initial distribution effects, with classical initial condition sampling techniques which ignore quantum dispersion and tunneling. Our results from all these various calculations will be compared with available experimental results enabling a detailed understanding of the reliability of the different approximations underlying these calculations.

\subsection{\label{sec:IC-sampling}Sampling the quantum initial density for the environment}
In this section we review the method developed by Poulsen and co-workers \cite{Poulsen03,Poulsen04,Poulsen05,Cao93,Cao94a,Cao94b,Cao94c,Cao94d,Cao94e,Cao90} that adapts the Feynman-Kleinert variational approach for obtaining an optimal local harmonic approximation to the Boltzmann operator which can be easily Wigner transformed. In our application to vibrational dephasing we suppose that the quantum vibration is prepared initially in some superposition state while the environmental degrees of freedom are unaffected by the vibrational excitation and are initially in thermal equilibrium with the vibrational degrees of freedom. As outlined in the previous section this quantity plays a central role as the phase space distribution function that must sampled to provide trajectory initial conditions in linearized approximations for the density matrix dynamics. We will explore the effects of the quantum nature of this initial equilibrium distribution on vibrational pure dephasing of the excited coherence. The development below in terms of the general Boltzmann operator is easily adapted to the case of the environmental degrees of freedom being in equilibrium with the quantum subsystem by supposing that the vibrational degree of freedom acts to provide an external force on the environment. As mentioned above we suppose that the experiments of interest can be interpreted as though the environment was initially in equilibrium with the ground state of the quantum subsystem so the Hamiltonian in the Boltzmann operator below describes interactions of the environment with the external field due to the ground state quantum subsystem.

The principle assumption underlying the Poulsen {\em et al.} \cite{Poulsen03,Poulsen04,Poulsen05} approach to generating an approximate Wigner transform of the Boltzmann operator, $\hat{\rho} = \exp[-\beta \hat{H}],$ is that the off-diagonal elements of the thermal density matrix $\rho_{\beta}(R,R')$ that enter the Wigner transform expression  
\begin{equation}
\rho^W_{\beta}(R,P) = \int d \Delta R \langle R + \Delta R/2 | \hat{\rho}_{\beta} | R - \Delta R/2 \rangle e^{-{i \over \hbar} P \Delta R}
\end{equation}
can be just as well represented as the diagonal elements that appear in the trace expression for the partition function. Accepting this proposal the approach proceeds as follows: First the Feynman-Kleinert method for computing the partition function is employed to obtain a local harmonic approximation to the Euclidian action that is variationally optimized for computing the trace of the density matrix. Next we assume that the off-diagonal elements can be approximated using the optimized local harmonic approximation and this form is employed to compute  the Wigner transformation. With this approach a local harmonic approximation is never directly made to the potential, rather an optimal local harmonic approximation to the Euclidean time action is found variationally using the partition function which involves integration over all space and thus contains global information. This local harmonic fitting procedure, however, is not influenced by any off-diagonal information. So it is unclear why it should provide a reliable approximation for generating the Wigner phase space density. One of our goals is to explore this question. In section \ref{sec:compare} we will directly compare density matrix elements for a simple model computed with in this approximation with exact results to explore the reliability of the off-diagonal density obtained with this and other approximate methods. 

As outlined above the derivation of the approach starts by following Feynman and Kleinert and considering the partition function, $Z$, at inverse temperature, $\beta = 1/k_B T$, for the environmental variables, $R$, written as a path integral over cyclic paths $(R(\tau): R(0) = R(\beta \hbar))$ {\em i.e.}
\begin{equation}
Z = \oint {\cal D}[R(\tau)] e^{-S[R(\tau)] /\hbar}
\end{equation} 
Where, $\oint {\cal D}[R(\tau)]$, means integrate over all points in the path, $R(\tau)$, including the common starting and finishing points, and the Euclidian action is $S[R(\tau)] = \int_0^{\beta \hbar} d \tau \{ {1\over 2} {\dot R}^T(\tau) {\bf M} {\dot R}(\tau) + V(R(\tau)) \}$. Here, ${\bf M}$, is the mass tensor, and $V(R)$, is the inter-particle interaction potential including the effects of external fields (such as, for example, coming from interactions with the vibrator in its ground state). In mass weighted cartesian coordinates, $q(\tau) = {\bf M}^{1/2} R(t)$, we can write 
\begin{equation}
Z = \oint {\cal D}[q(\tau)] e^{-{1\over \hbar}\int_0^{\beta \hbar} d \tau [ {1 \over 2} {\dot q}^2(\tau) + V(q(\tau))]}
\end{equation}
Since the paths of interest for $Z$ are real and periodic on $0 \le \tau \le \beta \hbar$ their Euclidean time dependence can be written in terms of a Fourier series {\em i.e.} 
\begin{equation}
q(\tau) = q_c + \sum_{n=1}^{\infty} [ q_n e^{i \Omega_n \tau} + q_n^* e^{-i \Omega_n \tau}] \label{eq:paths}
\end{equation}
where $\Omega_n = 2 \pi n / \beta \hbar$ are the Matsubara frequencies, the real zero frequency Fourier coefficient, $q_c = {1\over \beta \hbar} \int_0^{\beta \hbar} d\tau q(\tau)$, is the path centroid, and all other Fourier coefficients, $q_n$, are in general complex. 

If the configuration of the system is of dimension $d$ then the cyclic path space integration can be mapped onto an integration over Fourier coefficient vectors of length $d$ according to \cite{Kleinert86,Kleinert87,Roepstorff94,Kleinert04} 
\begin{equation}
\oint {\cal D}[q(\tau)] \rightarrow \int {dq_c \over (2 \pi \hbar^2 \beta)^{d/2}} \prod_{n=1}^{\infty} { \int d {\rm Re} q_n  \int d {\rm Im} q_n \over (\pi / \beta \Omega_n^2)^d}
\end{equation} 
and the partition function can be shown to be
\begin{eqnarray}
Z& =& \int {dq_c \over (2 \pi \hbar^2 \beta)^{d/2}} \prod_{n=1}^{\infty} { \int d {\rm Re} q_n  \int d {\rm Im} q_n \over (\pi / \beta \Omega_n^2)^d}\nonumber\\
& & \times  e^{-\beta \sum_{n=1}^{\infty} \Omega_n^2 |q_n|^2} e^{-{1 \over \hbar} \int_0^{\beta \hbar} d\tau V(q(\tau))}
\end{eqnarray}
Following Feynman and Kleinert \cite{Kleinert86,Kleinert87} the effective classical potential, $W(q_c)$, is defined as 
\begin{eqnarray} 
e^{-\beta W(q_c)} &=& \prod_{n=1}^{\infty} { \int d {\rm Re} q_n  \int d {\rm Im} q_n \over (\pi / \beta \Omega_n^2)^d}  \nonumber\\
& & \times e^{-\beta \sum_{n=1}^{\infty} \Omega_n^2 |q_n|^2} e^{-{1 \over \hbar} \int_0^{\beta \hbar} d\tau V(q(\tau))} \label{eq:veff}
\end{eqnarray}
and the quantum partition function is cast in the suggestive classical form 
\begin{equation}
Z=\int {dq_c \over (2 \pi \hbar^2 \beta)^{d/2}} e^{-\beta W(q_c)}
\end{equation}
identifying motion of the path centroid over the effective potential as a classical-like way to compute quantum properties using classical statistical mechanics ideas \cite{Kleinert86,Kleinert87,Feynman98}. Since $\beta \Omega_n^2 = 4 \pi^2 n^2 /\hbar^2 \beta$, at high temperature $(\lim_{\beta} \rightarrow 0)$, the Gaussian factors in the integrand of Eq.(\ref{eq:veff}), for example, will be dominated by small values of $|q_n|$, so under these conditions $q(\tau)$ will fluctuate only slightly from $q_c$ and, as Feynman and Kleinert suggested \cite{Kleinert86}, the effective potential might be computed using a local harmonic approximation to the full potential in the region around each path centroid location. They used the Gibbs-Bogoliobov-Feynman variational principle to determine the optimal local harmonic approximation at each centroid position. 

This approach starts from the exact expression for the partition function obtained by multiplying and dividing by some trial partition function $Z_{trial} = \oint {\cal D}[q(\tau)] e^{-{1 \over \hbar} S_{trial}[q(\tau)] }$ associated with some, yet to be specified, locally harmonic trial action, $S_{trial}$, thus 
\begin{eqnarray}
Z & = & {\oint {\cal D}[q(\tau)] e^{-{1 \over \hbar} \{ S[q(\tau)] - S_{trial}[q(\tau)] \} } e^{-{1\over \hbar} S_{trial}[q(\tau)] } \over \oint {\cal D}[q(\tau)] e^{-{1 \over \hbar} S_{trial}[q(\tau)] } } Z_{trial}\nonumber \\
\ & = & \left \langle e^{-{1 \over \hbar} \{ S[q(\tau)] - S_{trial}[q(\tau)] \} } \right \rangle_{trial} Z_{trial} 
\end{eqnarray}
Due to the concavity of the exponential function, $\langle e^{-f} \rangle \ge e^{-\langle f \rangle}$, so we replace the average of the exponential by the exponential of a more simply computed average and obtain the following variational result
\begin{equation}
 Z \ge e^{-{1 \over \hbar} \langle S[q(\tau)] - S_{trial}[q(\tau)] \rangle_{trial} } Z_{trial} \label{eq:var}
 \end{equation}
 If we choose $S_{trial}[q(\tau)] = \int_0^{\beta \hbar} d\tau \{ {1 \over 2} \dot{q}^2(\tau) + V_{trial}(q(\tau)) \}$ to be the Euclidian time action associated with motion in some trial potential $V_{trial}$, then 
\begin{eqnarray}
& &\langle S[q(\tau)] - S_{trial}[q(\tau)] \rangle_{trial} \nonumber\\
& &= \left \langle \int_0^{\beta \hbar}d\tau [V(q(\tau)) - V_{trial}(q(\tau))] \right \rangle_{trial} \label{eq:Sdiff}
\end{eqnarray}
Further since 
\begin{eqnarray}
& &\left \langle \int_0^{\beta \hbar} d\tau' f(q(\tau')) \right \rangle_{trial} \nonumber\\
& &= {1 \over Z_{trial}} \int_0^{\beta \hbar} d\tau' \oint {\cal D}[q(\tau)] f(q(\tau')) e^{-{1 \over \hbar} S_{trial}[q(\tau)]}
\end{eqnarray}
and the value of the cyclic path integral should be independent of the value of the time at which the function $f$ is evaluated since all $\tau'$ should be equivalent under the trace we can write \cite{Feynman-Hibbs65}, for example, 
\begin{equation}
\left \langle \int_0^{\beta \hbar} d\tau' f(q(\tau')) \right \rangle_{trial} = \beta \hbar \langle f(q(0)) \rangle_{trial} \label{eq:avint}
\end{equation}
To proceed we follow Feynman and Kleinert and suppose that the trial action is quadratic in displacements about the path centroid location thus 
\begin{eqnarray}
S_{trial}[q(\tau)]& =& \int_0^{\beta \hbar} d\tau {1 \over 2} [\dot{q}(\tau)^T \dot{q}(\tau) \nonumber\\
&+& (q(\tau) - q_c)^T{\bf K}(q_c)(q(\tau) - q_c) + L(q_c)]\nonumber\\
\end{eqnarray}
The term linear in displacement, $(q(\tau) - q_c)$, multiplying a derivative vector evaluated at $q_c$ vanishes identically when integrated over $\tau$ due to the cyclic nature of the paths on the interval $0 \le \tau \le \beta \hbar$ (Eq.(\ref{eq:paths})), so with this quadratic form only the offset term, $L(q_c)$, and the local curvature matrix, ${\bf K}(q_c)$, need to be determined variationally. 

Suppose the matrix, ${\bf U}(q_c)$, diagonalizes the curvature matrix giving a diagonal matrix of frequencies, ${\bf \omega}(q_c)$, according to ${\bf U}^T {\bf K} {\bf U} = {\bf \omega}^2$. Transforming to normal mode vectors, $\eta(\tau) = {\bf U}^T q(\tau) = \eta_c + \sum_{n=1}^{\infty} [\eta_n e^{i\Omega_n \tau} + \eta_n^* e^{-i \Omega_n \tau}]$, we find that the trial partition function is obtained as 
\begin{eqnarray}
Z_{trial}&=&  \int {d\eta_c \over (2 \pi \hbar^2 \beta)^{d/2}} \prod_{n=1}^{\infty} { \int d {\rm Re} \eta_n  \int d {\rm Im} \eta_n \over (\pi / \beta \Omega_n^2)^d} \nonumber\\
& &\times e^{-\beta \{ \sum_{j=1}^d \sum_{n=1}^{\infty} [\Omega_n^2 + \omega_j^2(\eta_c)]|\eta_{nj}|^2 + L(\eta_c) \}} \label{eq:Z-trial-normal}
\end{eqnarray}
The Gaussian integrals over the components of the $\eta_n$ can be performed analytically provided the frequencies $\omega_j$ have appropriate values (see detailed discussion below on the restrictions that this places on the matrix ${\bf K}$) and, using the fact that 
\begin{equation}
{\sinh x \over x} = \prod_{n=1}^{\infty} [1 + x^2/(n^2 \pi^2)]
\end{equation}
the trial partition function becomes
\begin{equation}
Z_{trial} = \int {dR_c \over (2 \pi \hbar^2 \beta)^{d/2}} e^{-\beta L(R_c)} \prod_{j=1}^d {\beta \hbar \omega_j(R_c)/2 \over \sinh \beta \hbar \omega_j(R_c)/2} \label{eq:ztrial}
\end{equation}
so in analogy to Eq.(\ref{eq:veff}) we can write
\begin{equation}
W_{trial}(R_c) = -{1 \over \beta} \sum_{j=1}^d \ln \left({\beta\hbar\omega_j(R_c)/2 \over \sinh \beta\hbar\omega_j(R_c)/2} \right) + L(R_c) \label{eq:Wtrial}
\end{equation}

The other quantities that need to be computed to apply the variational result in Eq.(\ref{eq:var}) according to Eqs.(\ref{eq:Sdiff}) - (\ref{eq:avint}), are averages of the potentials over the trial density. In Appendix \ref{app:avgpots} we outline the computation of these quantities obtaining the following general result written in terms of different Gaussian smeared potential forms: 
\begin{eqnarray}
\langle V(R) \rangle_{trial}& =& {1 \over Z_{trial}} \int {dR_c \over (2 \pi \hbar^2 \beta)^{d/2}} e^{-\beta L(R_c)} \nonumber\\
& &\times \prod_{j=1}^d {\beta \hbar \omega_j(R_c)/2 \over \sinh \beta \hbar \omega_j(R_c)/2} V_A(R_c) \label{eq:V-average}
\end{eqnarray}
where $V_A(R_c)$ is the Gaussian smeared full potential 
\begin{equation}
V_A(R_c) = \int_{-\infty}^{\infty} {dR \over |2\pi {\bf A}|^{1/2}} e^{-{1 \over 2} (R-R_c)^T {\bf A}^{-1}(R_c)(R-R_c)} V(R) \label{eq:VA-rspace-text}
\end{equation}
which can also be computed in Fourier space according to Eq.(\ref{eq:VA-kspace}). The Gaussian smeared local harmonic approximate trial potential is  
\begin{eqnarray}
& & V_A^{trial}(R_c) =  \int_{-\infty}^{\infty} {dR \over |2\pi {\bf A}|^{1/2}} e^{-{1 \over 2} (R-R_c)^T {\bf A}^{-1}(R_c)(R-R_c)} \nonumber \\
\ & \ & \times \left [{1 \over 2}(R-R_c)^T {\bf M}^{1/2} K(R_c) {\bf M}^{1/2}(R-R_c) + L(R_c) \right] \nonumber \\
& &= {1 \over 2} \sum_{ij} M_i^{1/2} K_{ij} M_j^{1/2} A_{ij} + L(R_c) 
\end{eqnarray}
which can be used in Eq.(\ref{eq:V-average}) to compute $\langle V_{trial}(R(0))\rangle_{trial}$.
In these results the smearing width matrix ${\bf A}$ is 
\begin{equation}
{\bf A} = {\bf M}^{-1/2} {\bf U} {\bf \Lambda} {\bf U}^T {\bf M}^{-1/2} \label{eq:A-def-text}
\end{equation}
with 
\begin{equation}
({\bf \Lambda})_{ij} = \sum_{n=1}^{\infty} {2 \over \beta[\Omega_n^2 + \omega_j^2(R_c)] } \delta_{ij} \label{eq:Lambda-def-text}
\end{equation}
as obtained in Appendix \ref{app:avgpots} (See Eq.(\ref{eq:A-def}) and Eq.(\ref{eq:Lambda-def})).

Thus we find the trial averaged action difference appearing in the variational expression, Eq.(\ref{eq:var}), can be written as 
\begin{eqnarray}
&& \langle S -S_{trial} \rangle_{trial} =  \beta \hbar \langle V(R(0)) - V_{trial}(R(0)) \rangle_{trial} \label{eq:Sdiff-fin} \nonumber \\
&& =  {1 \over Z_{trial}} \int {dR_c \over (2 \pi \hbar^2 \beta)^{d/2}} e^{-\beta L(R_c)} \prod_{j=1}^d {\beta \hbar \omega_j(R_c)/2 \over \sinh \beta \hbar \omega_j(R_c)/2} \nonumber \\
&& \times \beta \hbar[V_A(R_c) - {1 \over 2} \sum_{ij} M_i^{1/2} K_{ij} A_{ij} M_j^{1/2} - L(R_c)]
\end{eqnarray}

Using the above result we follow Feynman and Kleinert \cite{Kleinert86} and optimize the right hand side of Eq.(\ref{eq:var}), $f=\exp[-{1 \over \hbar}\langle S - S_{trial} \rangle_{trial}] Z_{trial}$,with respect to the variational parameter functions ${\bf K}(R_c)$ and $L(R_c)$. The details of this variational calculation are given in Appendix \ref{app:variational} and it gives the following relationships between the different optimized parameters:
\begin{equation}  
L(R_c) = V_A(R_c) - {1 \over 2} \sum_{ij} M_i^{1/2} K_{ij}(R_c) A_{ij}(R_c) M_j^{1/2} \label{eq:varL-text}
\end{equation}
Here the optimal curvature matrix is found from the gaussian smeared result
\begin{equation}
{\bf K}(R_c) = \int  {dR \over |2\pi {\bf A}|^{1/2}} {\bf D}(R) e^{-{1 \over 2} (R-R_c)^T {\bf A}^{-1}(R_c)(R-R_c)} \label{eq:K-fin-text}
\end{equation}
Where ${\bf D}(R)$ is the mass weighted Hessian 
\begin{equation}
{\bf D}(R) ={\bf M}^{-1/2} \left( {\partial^2 V \over \partial R \partial R^T}  \right)(R) {\bf M}^{-1/2}
\end{equation}

As outlined above the approach requires the computation of Gaussian smeared averages of various functions. The evaluation of such multidimensional integrals is in general complicated and could be performed with a MC procedure. In the application described here, however, the interaction potential functions being smeared are pair decomposable and, as outlined in Appendix \ref{app:smear-pair}, can be fit to Gaussian forms and the smearing integrals performed analytically.  

The structure of the above equations suggests the following iterative procedure for computing the smearing matrix ${\bf A}$: (1) Choose an initial guess for the ${\bf A}^{(0)}$ matrix. We use the previously converged smearing matrix for the last accepted centroid configuration. (2) Compute the full ${\bf K}^{(0)}$ matrix at the new centroid configuration using Eq.(\ref{eq:K-fin-text}) (or Eq.(\ref{eq:K-fin})). This can be done efficiently employing the symmetry of the local curvature matrix. (3) Diagonalize ${\bf K}^{(0)}$ to obtain the local harmonic frequencies $\omega^{(0)}(R_c)$ and the eigenvector matrix ${\bf U}^{(0)}$ as defined above Eq.(\ref{eq:Z-trial-normal}). Finally, (4), use these in Eqs.(\ref{eq:Lambda-def-text}) (or (\ref{eq:Lambda-def-1})) and (\ref{eq:A-def-text}) to obtain a new estimate of the smearing matrix, ${\bf A}^{(1)}$, then return to step (2) and iterate this process till convergence. 

Using the converged set of parameters, we compute the various Gaussian smeared quantities (using Eq.(\ref{eq:VA-rspace-text}) for example) and obtain the variationally optimized value for the energy offset parameter $L(R_c)$ as in Eq.(\ref{eq:varL-text}) at the new centroid configuration.  

The final steps in the approach involve writing the density operator approximately as the integral over centroid locations of the variationally optimized local harmonic form around each centroid position and then analytically Wigner transforming this locally harmonic result. Detail of these calculations are given in Appendix \ref{app:rhoH-WT}. The approximate density has the form 
\begin{widetext}
\begin{eqnarray}
e_{trial}^{-\beta {\hat H}} & = & \int d\eta' \int d\eta'' \int \frac{d\eta_c}{(2 \pi)^d}
e^{-\beta W_{trial}(\eta_c)} |\eta' \rangle \langle \eta''| \nonumber \\
\ & \ & \times \prod_i [ \exp \left\{ -\frac{\omega_i}{\hbar \alpha_i} (\frac{\eta_i'' + \eta_i'}{2} - \eta_{ci})^2 - \frac{\omega_i}{4 \hbar} \coth(\beta \hbar \omega_i / 2) (\eta_i'' - \eta_i')^2 \right\}
{\sqrt{\frac{\omega_i}{ \pi \hbar \alpha_i}}} {\sqrt{\frac{2 \pi}{\hbar^2 \beta}}}]
\label{eq:Boltz-fin-text}
\end{eqnarray}
Here we define the quantity 
$\alpha_i = \coth \beta \hbar \omega_i/2 - 2/\beta \hbar \omega_i$
and Wigner transformation yields
\begin{eqnarray}
(e_{trial}^{-\beta {\hat H}})_W(Q,P_Q) & = & \int \frac{d\eta_c}{(2 \pi \hbar)^d} e^{-\beta W_{trial}(\eta_c)}
\prod_{i=1}^d \{ \left(\frac{4\pi}{\coth(\beta \hbar \omega_i/2) \alpha_i \beta/2} \right)^{1/2} \label{eq:Boltz-Wigner-text}\\
\ & \ & \times \exp[-\frac{\omega_i(\eta_c)}{\alpha_i \hbar} (Q_i-\eta_{ci})^2] \exp[-\frac{\tanh(\beta\hbar\omega_i/2)}{\hbar \omega_i} P_{Q_i}^2] \} \nonumber
\end{eqnarray} 
\end{widetext}
The integral in Eq.(\ref{eq:Boltz-Wigner-text}) is easily computed by importance sampled Monte Carlo. Thus a Cartesian centroid configuration, $R_c$, is sampled. The procedure outlined above Eq.(\ref{eq:Boltz-fin-text}) is employed to iterate the width, ${\bf A}^{(n)}$, and curvature, ${\bf K}^{(n)}$, matrices to convergence. As indicated, this involves diagonalizing ${\bf K}^{(n)}$ at each iteration to obtain the local harmonic frequencies and normal modes. The converged value of $W_{trial}$ computed from Eq.(\ref{eq:Wtrial}) is used to accept or reject the sampled centroid configuration. Next, to sample the $(Q,P_Q)$ phase space distribution associated with the approximate Wigner transformed Boltzmann operator, a set of $k=1,M$ normal mode displacements with components $\xi_i^{(k)} = Q_i^{(k)} - \eta_{ci}$ are sampled from the component Gaussian distributions with variances $\sigma_{Qi}=(\alpha_i \hbar / 2 \omega_i)^{1/2}$ and the set of $M$ points in Cartesian space $R^{(k)}=R_c + {\bf M}^{-1/2} {\bf U}(R_c) {\bf \xi}^{(k)}$ is generated to provide a sampling of initial configurations around the point $R_c$. Similarly, Gaussian random numbers $\zeta_i^{(k)}$ are sampled with variance $\sigma_{Pi} = (\hbar \omega_i / 2 \tanh(\beta \hbar \omega_i/2))^{1/2}$ to provide a set of normal mode momentum vectors  $P_{Q_i}^{(k)}$ which are transformed to cartesian initial momenta according to $P^{(k)} = {\bf M}^{1/2} {\bf U}(R_c) \zeta^{(k)}$. The numerical factor $(\dots)^{1/2}$ in Eq.(\ref{eq:Boltz-Wigner-text}) provides an automatic centroid configuration dependent normalization for the sampling of these Gaussian distributions, and so each sample phase space configuration generated in this way carries unit weight.

\subsection{\label{sec:compare}Comparison of Initial Condition Sampling Methods for Model System}
To test the reliability of our implementation of the Feynman-Kleinert approach for sampling the Wigner distribution outlined in the previous subsection we have applied it to several simple exactly solvable one dimensional model systems. In this section we present these results and compare with exact calculations and with another approximate approach due to Shi and Geva (SG) \cite{ShiGeva03a,ShiGeva03b,ShiGeva03c}. These authors proceed by multiplying and dividing the Wigner transform expression of the Boltzmann operator by the diagonal elements of the thermal density operator, thus
\begin{eqnarray}
(e^{-\beta\hat{H}})_W(Q,P)&=&\langle Q|e^{-\beta\hat{H}}|Q \rangle\int dz e^{-\imath P z/\hbar}\nonumber\\
&\times & \frac{\langle Q-z/2|e^{-\beta\hat{H}}|Q+z/2\rangle}{\langle Q|e^{-\beta\hat{H}}|Q \rangle}
\end{eqnarray}
They make the Local Harmonic Approximation (LHA) to the potential, expanding to quadratic order about the position $Q$. Note this approximation is made only in the ratio of off-diagonal to diagonal density matrix elements in the integrand, however, the full anharmonic dependence  of prefactor diagonal element is included through path integral calculations. With this local harmonic form, the Gaussian integrals can be performed analytically yielding the following result for the case of one dimension:
\begin{eqnarray}
(e^{-\beta\hat{H}})_W^{SG}(Q,P)&=&\langle Q|e^{-\beta\hat{H}}|Q \rangle (\frac{4\pi \hbar }{M\omega(Q)\chi(Q)})^{1/2}\nonumber\\
& &\times\exp[-\frac{P^2}{\hbar M \omega(Q) \chi(Q)}]
\end{eqnarray}
Where $\chi(Q)$ depends on the local curvature of the potential about the point $Q$ and has the form
$\chi(Q)=\coth\frac{\beta\hbar\omega(Q)}{2}$ with $\omega(Q)=[\partial^2 V(Q)/\partial Q^2/M]^{1/2}$.   

For the purpose of comparison with exact results we computed the density matrix for our 1D model using the numerical matrix multiplication (NMM) approach \cite{Storer73,Thirumalai83}. It is most convenient to make direct comparison with the full density matrix. The approximate Wigner densities can be inverse Wigner transformed to give approximate coordinate space density matrices using to the following result:
\begin{eqnarray}
&&\langle Q|e^{-\beta\hat{H}}|Q' \rangle=(2\pi\hbar)^{-1}\int dP \nonumber\\
&&\times  e^{-\imath P(Q-Q')/\hbar} (e^{-\beta\hat{H}})_W(\frac{1}{2}(Q+Q'),P)
\end{eqnarray}
Thus the SG approximation to the density matrix becomes
\begin{equation}
\langle Q|e^{-\beta\hat{H}}|Q'\rangle^{SG}=\langle \bar{Q}|e^{-\beta\hat{H}}|\bar{Q} \rangle \exp[-\frac{M\omega(\bar{Q})\chi(\bar{Q})}{4\hbar}(Q-Q')^2] \label{eq:rhoSG}
\end{equation}
with  $\bar{Q} = (Q+Q')/2$. The corresponding FK approximation to the density matrix takes the form:

\begin{eqnarray}
&&\langle Q |e^{-\beta\hat{H}}|Q'\rangle^{FK} = (2\pi\hbar)^{-1} \int dQ_c e^{-\beta W(Q_c)} \nonumber \\
&&\times 
\left(\frac{2M^2\omega(Q_c)}{\beta\hbar \alpha}\right)^{1/2} \exp[-\frac{M \omega(Q_c)}{\hbar \alpha}(\bar{Q}-Q_c)^2] \nonumber \\
&& \times  \exp[-\frac{M\omega(Q_c) [\alpha(Q_c)+2/\beta\hbar\omega(Q_c)]}{4\hbar}(Q-Q')^2] \label{eq:rhoFK}
\end{eqnarray}

The FK form in Eq.(\ref{eq:rhoFK}) and the SG approximate density matrix of Eq.(\ref{eq:rhoSG}) share some similar features. Setting $Q=Q'$ in Eq.(\ref{eq:rhoFK}) gives an expression for the diagonal elements $\langle \bar{Q} | e^{-\beta \hat{H}} | \bar{Q} \rangle^{FK}$. If we evaluate the gaussian in $(Q-Q')$ of Eq.(\ref{eq:rhoFK}) at the maximum of the first gaussian in this equation {\em i.e.} $Q_c=\bar{Q}$ in an effort to approximate the $Q_c$ integration, we recover a form similar to the SG approximation of Eq.(\ref{eq:rhoSG}). In general however the full integral over the centroid position must be performed so we expect differences between these results. The most significant differences must be due to the fact that the SG form employs a single local harmonic frequency $\omega(\bar{Q})$ to compute the off-diagonal elements. The FK expression, on the other hand, combines results from a range of smeared frequencies that are {\em not} obtained by local harmonic approximation to the bare potential. 

\begin{figure*}
\centerline{\includegraphics[width=2.5in]{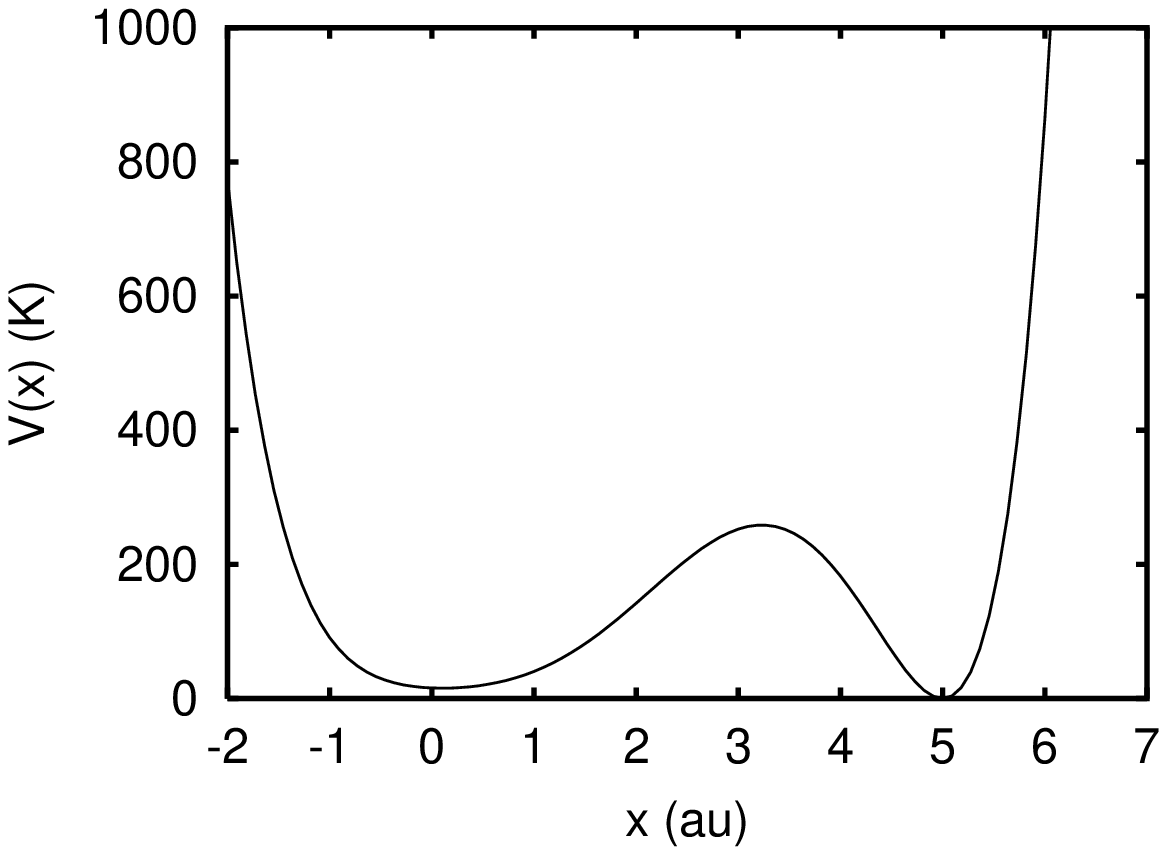}\includegraphics[width=3.in]{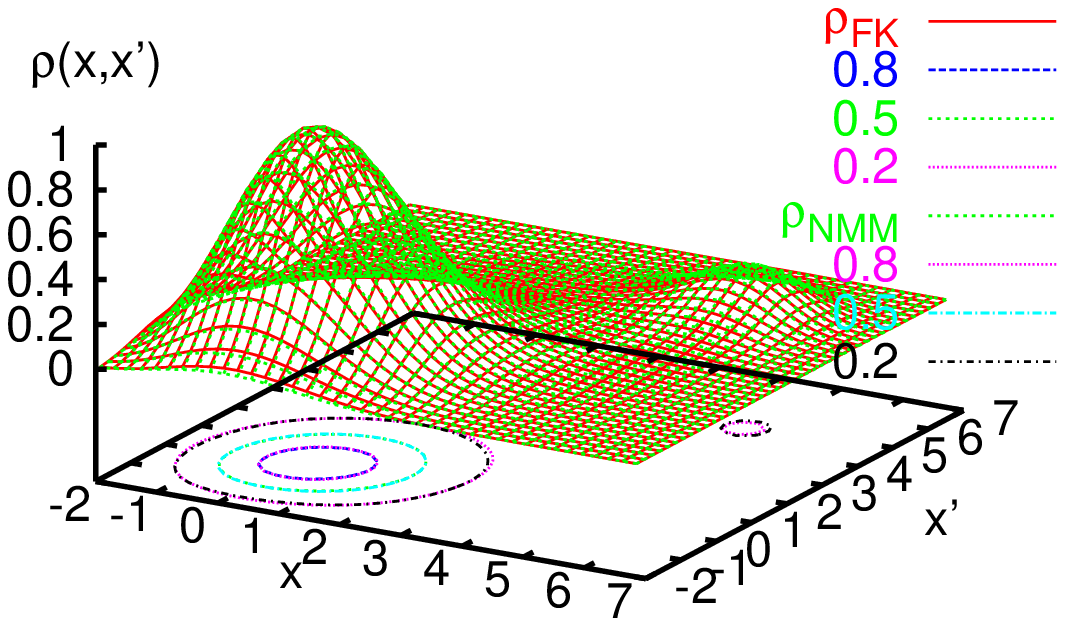}}
\caption{\label{fig:pot-FKNMM}Left panel: Veneziano asymmetric double well potential. Right panel: Asymmetric double well density matrix, green surface is exact numerical calculation and red surface are results from FK approximation.}
\end{figure*}

\begin{figure*}
\centerline{\includegraphics[width=3.in]{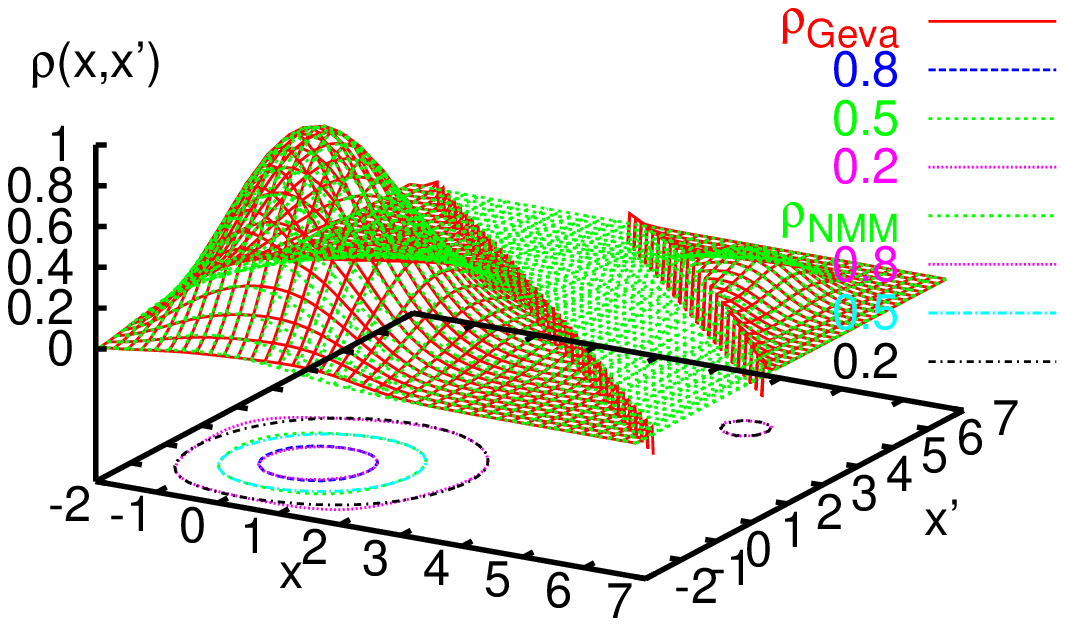} \includegraphics[width=3.in]{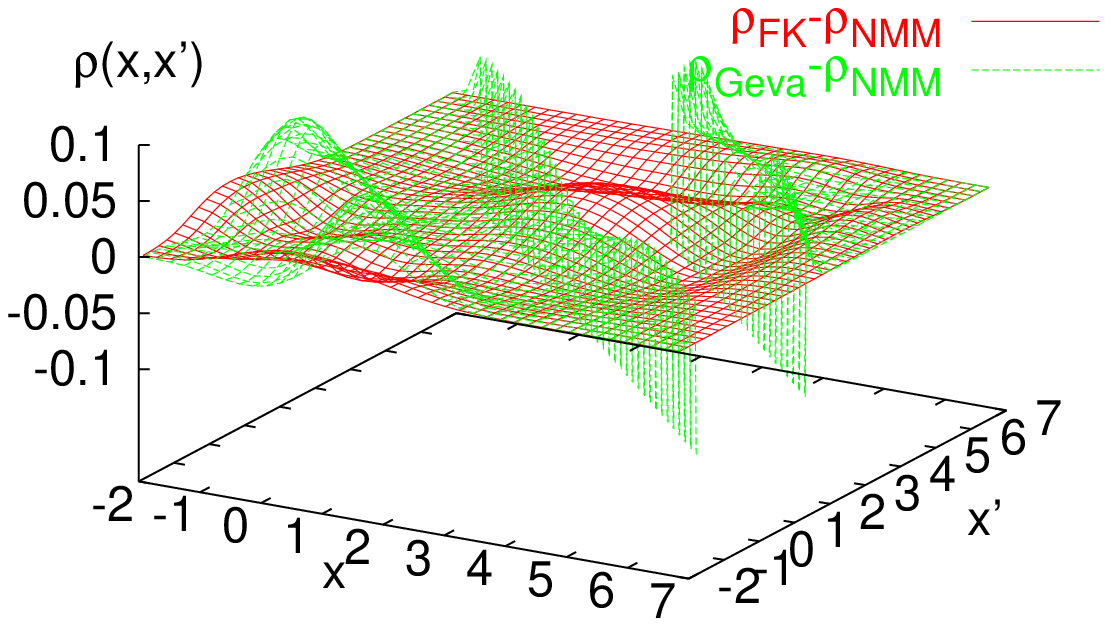}}
\caption{\label{fig:SGNMM-Diff}Left panel: Asymmetric double well density matrix, green surface is exact numerical calculation and red surface are results from SG approximation. Right panel: Red surface gives difference between exact density matrix and FK approximation. Green surface presents differences between exact density matrix and SG approximation.}
\end{figure*}

These differences are apparent in the results presented in Figures. \ref{fig:pot-FKNMM}, and \ref{fig:SGNMM-Diff}.  In these calculations we explored the approximate density matrixes for an asymmetric double well potential (the Veneziano potential\cite{Karsch84}) of the form $V(Q) = {1 \over 2} E_c (1+Q^2)^2(1-mx)^2$ with the following parameter values in atomic units: $E_c = 1\times 10^{-4}$, $m=0.2$ and $M=1600$. Results are presented with a temperature corresponding T=50K.
 
From Fig. \ref{fig:pot-FKNMM} it is clear that under these conditions the FK approximation reproduces the density matrix very accurately across the entire region, even when barrier tunneling is important. In Fig. \ref{fig:SGNMM-Diff}, on the other hand, we see that the SG approach gives undefined results for the density matrix in the tunneling region. Moreover, the density matrix difference results presented in the right hand panel show that even though the SG density matrix is exact along the diagonal (when it can be defined) the off diagonal elements differ considerably from the exact results compared to the significantly smaller differences observed with the FK approximation.  

The general condition for convergence of the various multidimensional Gaussian integrals like Eq.(\ref{eq:VA-kspace}) in ${\bf k}$-space, or analogously Eq.(\ref{eq:VA-rspace-text}) in ${\bf R}$-space is that the gaussian width matrix, ${\bf A}$, appearing in these expressions should be positive definite, {\em i.e.}, all its eigenvalues should be positive.  From the definition of ${\bf A}$ in Eq.(\ref{eq:A-def-text}) the eigenvalues are proportional to $\Lambda_{ll}$  as given in Eq.(\ref{eq:Lambda-def-text}) (or Eq.(\ref{eq:Lambda-def-1})). The right hand side of this expression is positive provided all the eigenvalues of ${\bf K}$ satisfy $\omega_l^2 > -(2 \pi /\beta \hbar)^2=-K_m^{FK}$ {i.e.} some negative frequencies can be tolerated with in the FK approach and this is an important factor in reproducing the density matrix in the tunneling region. Analysis of the SG density matrix expression in Eq.(\ref{eq:rhoSG}) reveals that it too can tolerate negative curvature regions. It is found that $\omega^2 > -(\pi /\beta \hbar)^2=-K_m^{SG}$ \cite{ShiGeva03b}. Thus the maximum negative curvatures $K_m^{FK}$ that the FK approach can tolerate are 4 times those of the SG approach. Moreover, due to the use of the smeared potential in the FK approach as distinct from the bare potential in the SG formulation, curvatures with the FK approach are generally significantly smaller in absolute value than those of the SG approach. Thus the FK approach is expected to have a considerably wider range of applicability.

\section{\label{sec:Results}Results}

\subsection{Outline of Experiments}
The TR-CARS experiments of Apkarian and co-workers \cite{Apkarian00,Apkarian01a,Apkarian01b,Apkarian04a,Apkarian04b,Apkarian04c,Apkarian05a,Apkarian05b,Apkarian05c} involve exciting controllable coherent superpositions of I$_2$ vibrational states in rare gas matrices. After the excitation pulses that prepare the initial vibrational superposition on a chosen electronic state, probing pulses project the evolving packet onto other electronic states and the time dependence of the emission from these states gives a signal that can be related to the evolution of the initial coherence. Apkarian and co-workers use a model of these experiments to extract dephasing rates of the different vibrational superposition states they prepare. Their work exploring these dephasing rates for vibrational superpositions prepared in the ground $X$ electronic state is the focus of our studies here. In our work we assume that the pure dephasing rate can be obtained from the long time exponential decay rate of the off-diagonal elements of the density matrix in an appropriately chosen vibrational representation. Martens and co-workers \cite{Martens04,Martens05a,Martens05b,Martens06} have used their semi-classical Liouville dynamics approach (which is formally equivalent to the linearized dynamics employed here) to study this problem and have presented a useful model of the relevant vibrational state dependent interactions. We follow these workers and assume that the eigenstates of a Morse oscillator whose parameters are fit to give solution phase experimental vibrational data for the ground $X$ electronic state provides such an appropriate vibrational representation. Using this model of the interactions we explore the linearized approach for computing the dynamics of the vibrational density matrix with different distributions of initial conditions. The main difference between our calculations and this previous work is the use of the FK-Wigner quantum initial condition sampling approach developed in the previous section.  

\subsection{Computational Model}
The computational model employed in our studies of vibrational dephasing is closely related to that of Martens and co-workers \cite{Martens04,Martens05a,Martens05b,Martens06} and Meier and Beswick \cite{Beswick04a,Beswick04b}. 
Thus we write the Hamiltonian as $\hat{H}=\hat{H}_s + \hat{H}_b + \hat{H}_{s-b}$ where 
\begin{equation}
\hat{H}_s = {\hat{p}_r \over 2 \mu} + V_s(\hat{r}) 
\end{equation}
\begin{equation}
\hat{H}_b = {\hat{p}_{com}^2 \over 2 m} + {\hat{P}^2 \over 2M} + V_b(\hat{Q})
\end{equation}
\begin{equation}
\hat{H}_{s-b} = {\hat{L}^2 \over 2 \mu \hat{r}^2} + V_{s-b}(\hat{r}, \hat{\theta}, \hat{\phi}, \hat{r}_{com}, \hat{Q}) 
\end{equation}
Here $\mu$ is the reduced mass of the diatomic whose position is specified by $\hat{r}, \hat{\theta}, \hat{\phi}, \hat{r}_{com}$ and $\hat{Q}$ describes the configuration of the bath. In our calculations we represent the quantum vibration in terms of a basis set of eigen states of system Hamiltonian, {\em i.e.} $\hat{H}_s |v \rangle = \epsilon_v |v \rangle$. We employ the approach of Martens and co-workers \cite{Martens05a,Martens05b,Martens06} and approximate the density associated with these vibrational basis states using two weighted $\delta$-functions. Thus we write the vibrational wave functions density as $|\langle r|v\rangle|^2 = \sum_{k=1}^2 c_k^{vv} \delta(r-r_k^{vv})$ and the $\delta$-function positions $r_k^{vv}$ and weights $ c_k^{vv}$, which are state dependent, are fit to gas phase vibrational bond length moments \cite{Herzberg50}. Here we explore vibrational pure dephasing so  $h_{\alpha \beta}$ for $\alpha \ne \beta$ are assumed to be small and the full Hamiltonian becomes 
\begin{equation}
\hat{H} = {\hat{p}_{com}^2 \over 2 m} + {\hat{P}^2 \over 2M} + \sum_{\alpha} |\alpha \rangle h_{\alpha \alpha} (\hat{r}_{com},\hat{\theta},\hat{\phi}, \hat{L}, \hat{Q}) \langle \alpha|
\end{equation}
where the diagonal Hamiltonian matrix elements are obtained by summing interactions of the different bond length molecular representations with the environment as 
\begin{eqnarray}
&&h_{\alpha \alpha} (\hat{r}_{com},\hat{\theta},\hat{\phi}, \hat{L}, \hat{Q}) = \epsilon_v + V_b(Q)\nonumber\\
&& + \left(p_{\theta}^2 + {p_{\phi}^2 \over \sin^2 \theta} \right) {1 \over 2 \mu \sum_k c_k^{vv} {r_k^{vv}}^2}\nonumber\\
&& + \sum_k c_k^{vv} V_{s-b}(r_k^{vv}, \theta, \phi, r_{com}, Q) \label{eq:Ham}
\end{eqnarray}
In our calculations we employed the Morse potential model of Martens and coworkers \cite{Martens05a,Martens05b,Martens06} to describe the I$_2$ vibrator and the interactions between the iodine atoms and the krypton particles as well as the solvent-solvent interactions are modeled using the Lennard-Jones potentials of these workers. 

This Hamiltonian model is incorporated in Eq.(\ref{eq:Lin-den}) which gives that the reduced vibrational density matrix elements can be computed by averaging the basis state energy  gap phase factor along a classical trajectory in which the environment moves on the mean potential surface produced by the two states involved in the coherence. In our studies the initial conditions for the environmental variables are sampled from either the classical distribution (as used in Marten's earlier work) or the FK approximation to the Wigner transform of the bath density matrix. In our implementation of this quantum initial condition sampling the rotor is fixed at the minimum energy orientation and center of mass position for the ground vibrational state and optimized solvent geometry. The solvent particle positions and momenta are then sampled from the approximate Wigner transform in the presence of the fixed rotor. This approach thus neglects the quantum rotational dispersion. The rotational dynamics, however is incorporated, as the system evolves on the mean potential surface using rigid rotor MD \cite{Allen87} with the weighted moment of inertia appearing in the second last term in Eq.(\ref{eq:Ham}).

In our studies the iodine molecule replaces two krypton atoms in a double substitutional site in a 108 particle FCC lattice. 

In implementing the FK sampling approach of Eq.(\ref{eq:Boltz-Wigner-text}) we perform regular Metropolis MC sampling of the centroid position and accept configurations based on the change in $W_{trial}$. The solution of the variational equations was generally found to converge in a very small number of iterations, typically two or three. For each sampled centroid position we use gaussian sampling to generate 5 phase space points according to the FK approximate Wigner distribution which serve as trajectory initial conditions. All the results presented below have been averaged over 5000 independent trajectories.

\subsection{Dephasing Time Results}
\begin{figure}
\centerline{\includegraphics[width=4.in]{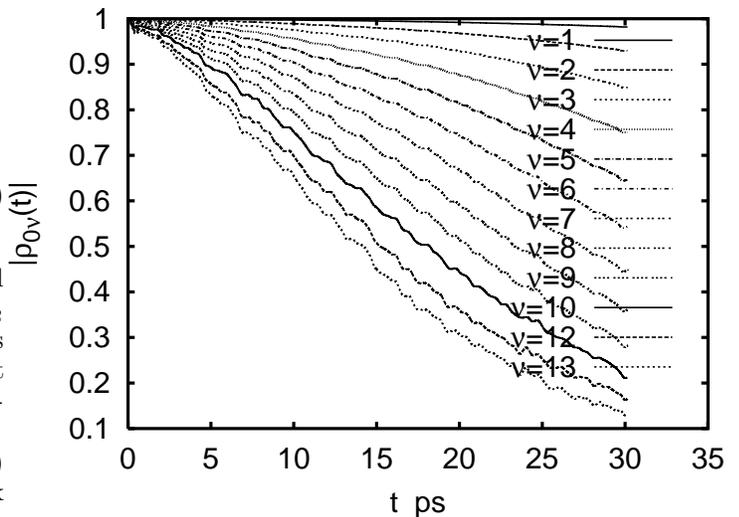}}
\caption{\label{fig:2_6rhovt}Decay of various ground - excited state elements of the reduced vibrational density matrix of I$_2$ in for solid krypton at T=2.6K obtained with FK Wigner initial condition sampling.}
\end{figure}

In Fig. \ref{fig:2_6rhovt} we present results showing the time dependence of the off-diagonal elements (ground and excited state) of the reduced vibrational density matrix. As expected the coherence decays extremely slowly for excitation of low energy superposition states. However, due to the disparate nature of the interactions between the ground state molecule with its environment and a highly excited molecule and its environment, the large fluctuations in energy gap between these states, in this situation, cause rapid decay of these off-diagonal density matrix elements.  

\begin{figure*}
\centerline{\includegraphics[width=2.5in]{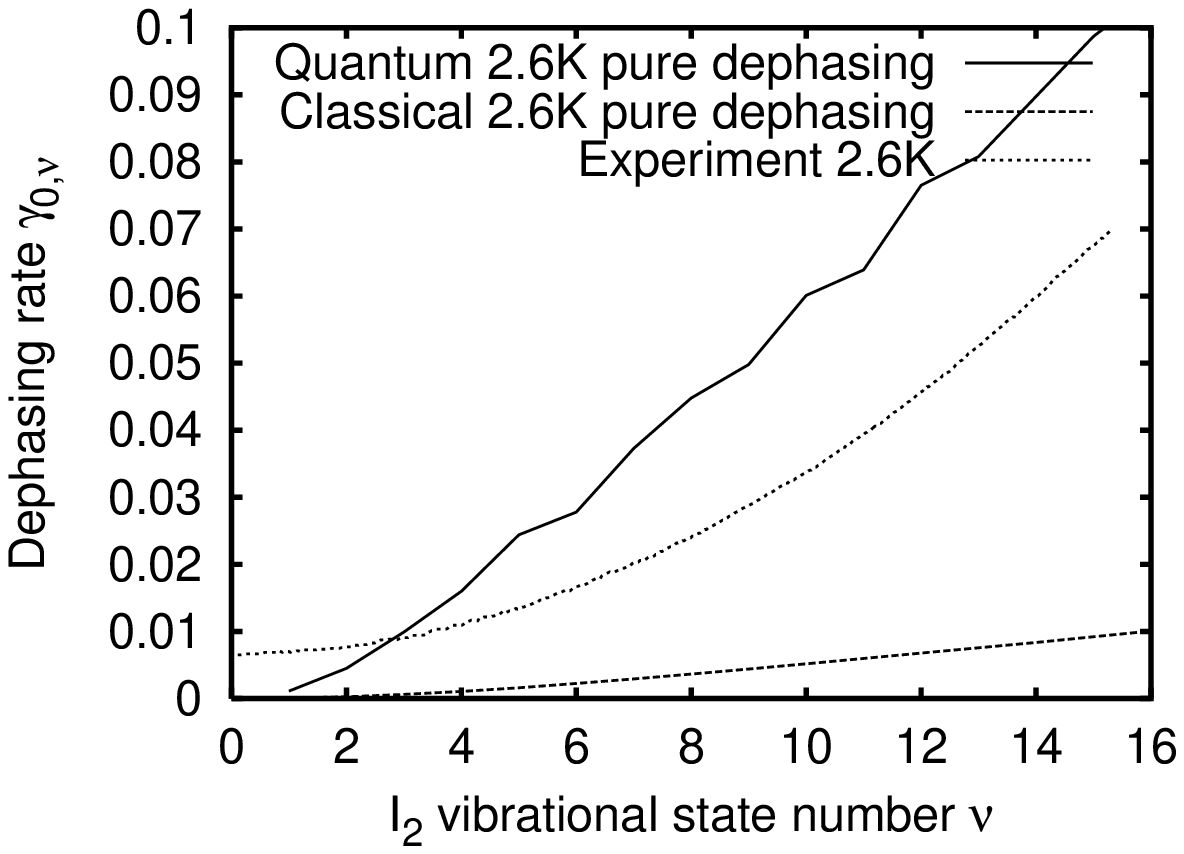} \includegraphics[width=2.5in]{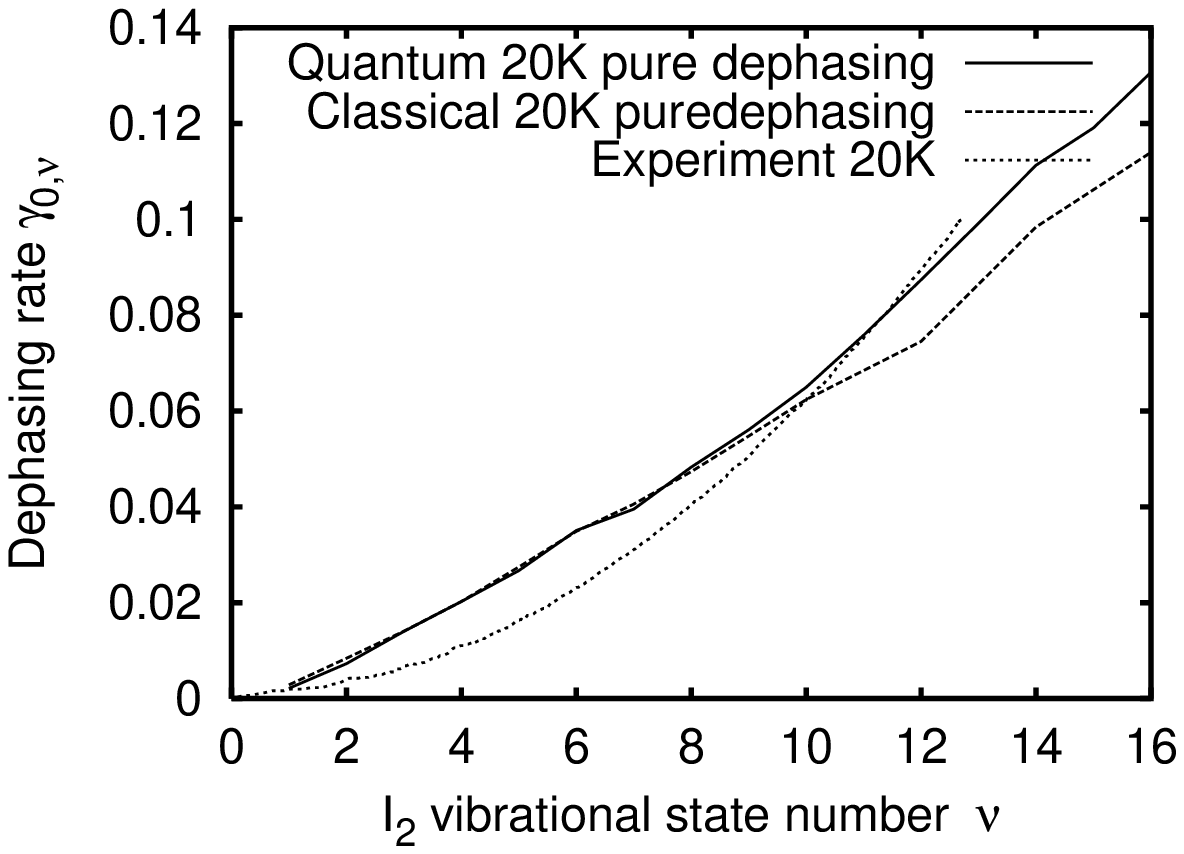}
\includegraphics[width=2.5in]{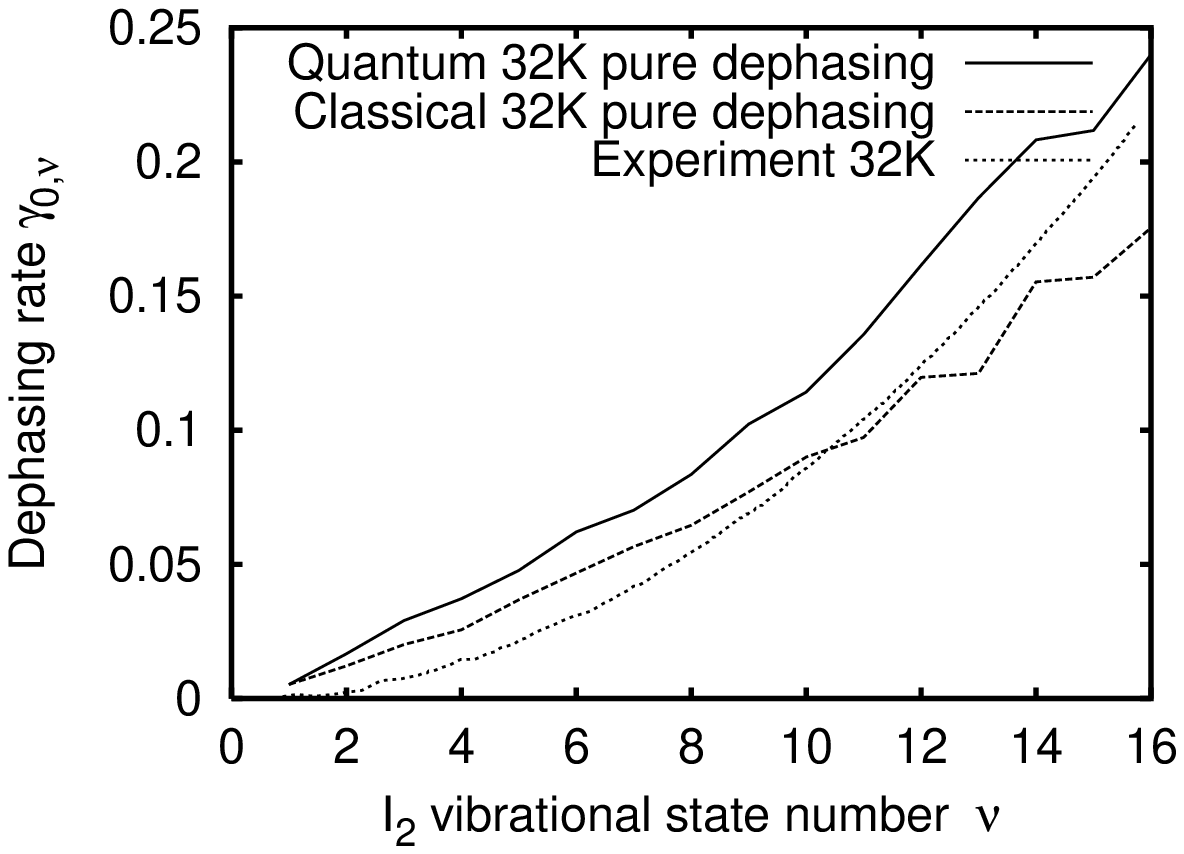}}
\caption{\label{fig:gammas}Comparison of experimental and calculated vibrational pure dephasing rates as functions of vibrational quantum state. Each panel compares FK-Wigner and Classical initial condition sampling calculations with experimental results. The different panels report results for various temperatures: left panel, T=2.6K; center panel, T=20K; right panel, T=32K}
\end{figure*}

We have extracted the dephasing rates from our linearized dynamics calculations of the reduced density matrix by fitting the long time behavior of the data in Fig. \ref{fig:2_6rhovt} to an exponential decay form. The computed dephasing rates, for classical and FK-Wigner initial condition sampling calculations are compared with the experimental results of Apkarian and co-workers \cite{Apkarian05a,Apkarian05b} in Fig. \ref{fig:gammas}. Here it is clear that at the higher temperatures, T=20K and 32K, (center and right hand panels) the nature of the initial condition sampling has little effect on the calculated dephasing rate and that these calculation results agree well with the experimental results. The fact that the results from FK-Wigner and classical initial condition sampling agree reasonably well at high temperatures suggests that our implementation of the FK-Wigner sampling approach is reliable. 

At low temperatures, for example T=2.6K, however, (see left panel in Fig. \ref{fig:gammas}) the approach used to sample initial conditions has a significant influence on the computed dephasing rates. The rates obtained with classical initial conditions are almost and order of magnitude smaller than the experimental dephasing rates observed at this low temperature. Dephasing rates obtained with the FK-Wigner sampling approach, however, are on the same order as experimental results and show a very similar increase in dephasing rate with vibrational state. As discussed in detail by Apkarian and co-workers \cite{Apkarian05a,Apkarian05b}, these low temperature experimental results presented in Fig. \ref{fig:gammas} are influenced by strain induced inhomogeneity of the trapping sites. They suggest that their results at the low quantum numbers are thus sensitive to sample preparation at these low temperatures. The fact that the slope of the experimental curve agrees well with our computed dephasing rates when FK-Wigner initial condition sampling is employed suggests that this approach is a reasonably accurate way of incorporating quantum dispersion effects in linearized path integral dynamics calculations. While showing good qualitative behavior, generally the dephasing rates obtained with FK-Wigner sampling are about a factor of 2 too fast. This discrepancy could arise from two possible sources: First the use of the FK variational approach for approximating the Wigner transform means that the trial harmonic free energy is always larger than the true free energy. Thus the approach overestimates the effects of quantum fluctuations and dispersion which is essentially equivalent to a higher effective temperature and could result in too rapid dephasing. The second source of deviation is the semiclassical nature of the linearized dynamics in which classical trajectories are evolved over the mean surface from the approximate initial condition distribution. 

In the next section we explore the reasons behind the difference between classical and FK-Wigner dephasing rates by comparing structural differences in the simulated environments.

\subsection{Equilibrium Structure}
\begin{figure*}
\centerline{\includegraphics[width=3.in]{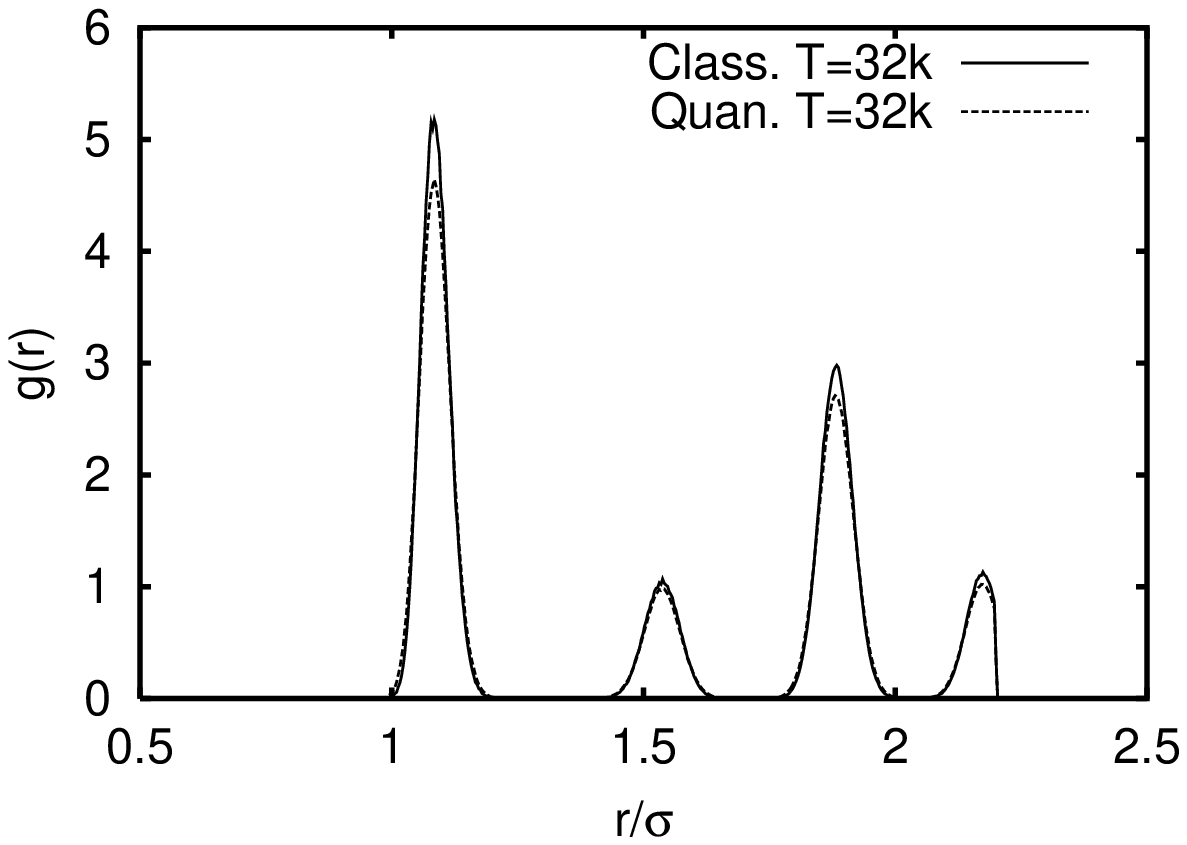} \includegraphics[width=3.in]{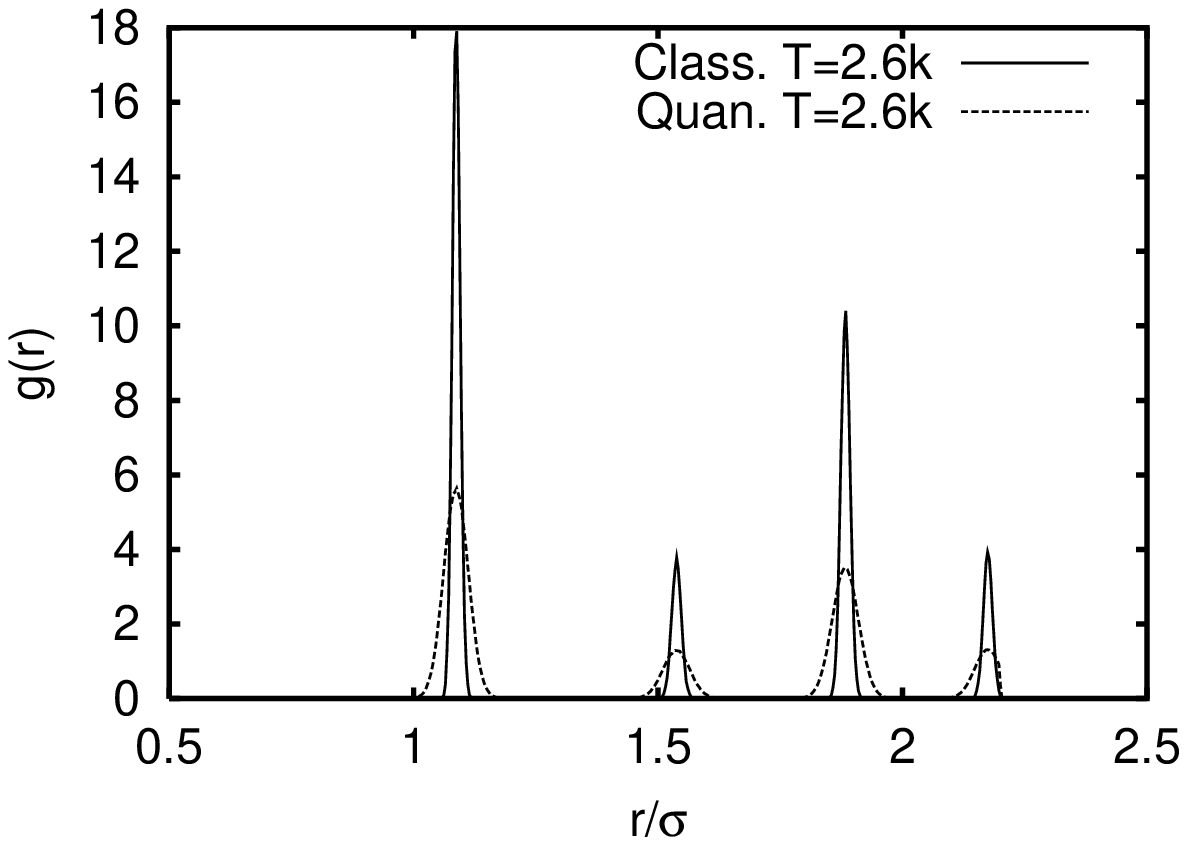}}
\caption{\label{fig:gr}Left panel compares pure solvent (solid krypton) pair distribution functions at T=32K computed using Monte Carlo sampling with the Classical Boltzmann distribution (solid curve), and the FK approximation to the Wigner transformed equilibrium distribution (dashed curve). Right panel presents the same results but for T=2.6K.}
\end{figure*}

In Fig. \ref{fig:gr} we compare pair distributions computed with FK-Wigner equilibrium distribution sampling using Eq.(\ref{eq:Boltz-Wigner-text}) and results from classical Boltzmann distribution sampling in a pure Krypton crystal. At T=32K the classical and quantum pair distributions agree very closely, suggesting that the temperature is sufficiently high under these conditions that quantum effects are unimportant. These distributions provide initial conditions for the dynamics underlying the dephasing rate calculations. Thus the similarities in structure manifest themselves in similar dephasing rates as presented in the right panel of Fig. \ref{fig:gammas}. At T=2.6K, however, the classical pair distribution function is strongly peaked compared to the FK-Wigner result (see right panel in Fig. \ref{fig:gr}) so the distribution of trajectory energy gap fluctuations resulting from classical sampling will be narrow, leading to slow dephasing compared to the FK-Wigner result as discussed in connection with Fig. \ref{fig:gammas}. 

\begin{figure}
\centerline{\includegraphics[width=4.in]{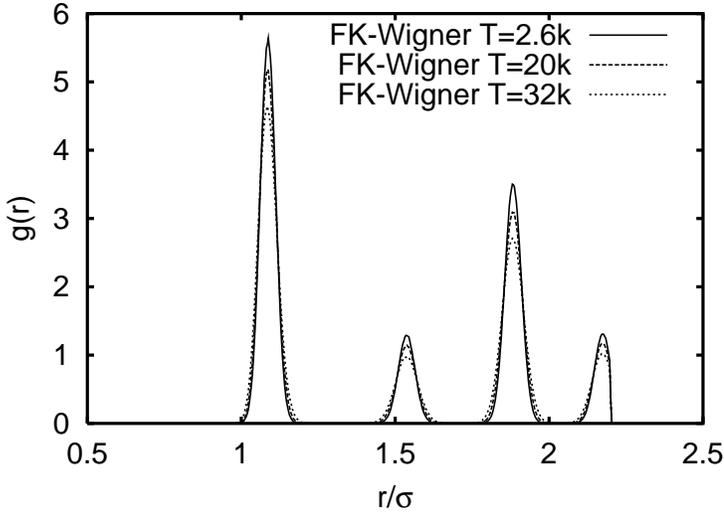}}
\caption{\label{fig:grQ}Pair distribution functions for solid krypton at various temperatures computed use FK-Wigner approximate quantum equilibrium distribution.}
\end{figure}

\begin{figure*}
\centerline{\includegraphics[width=3.in]{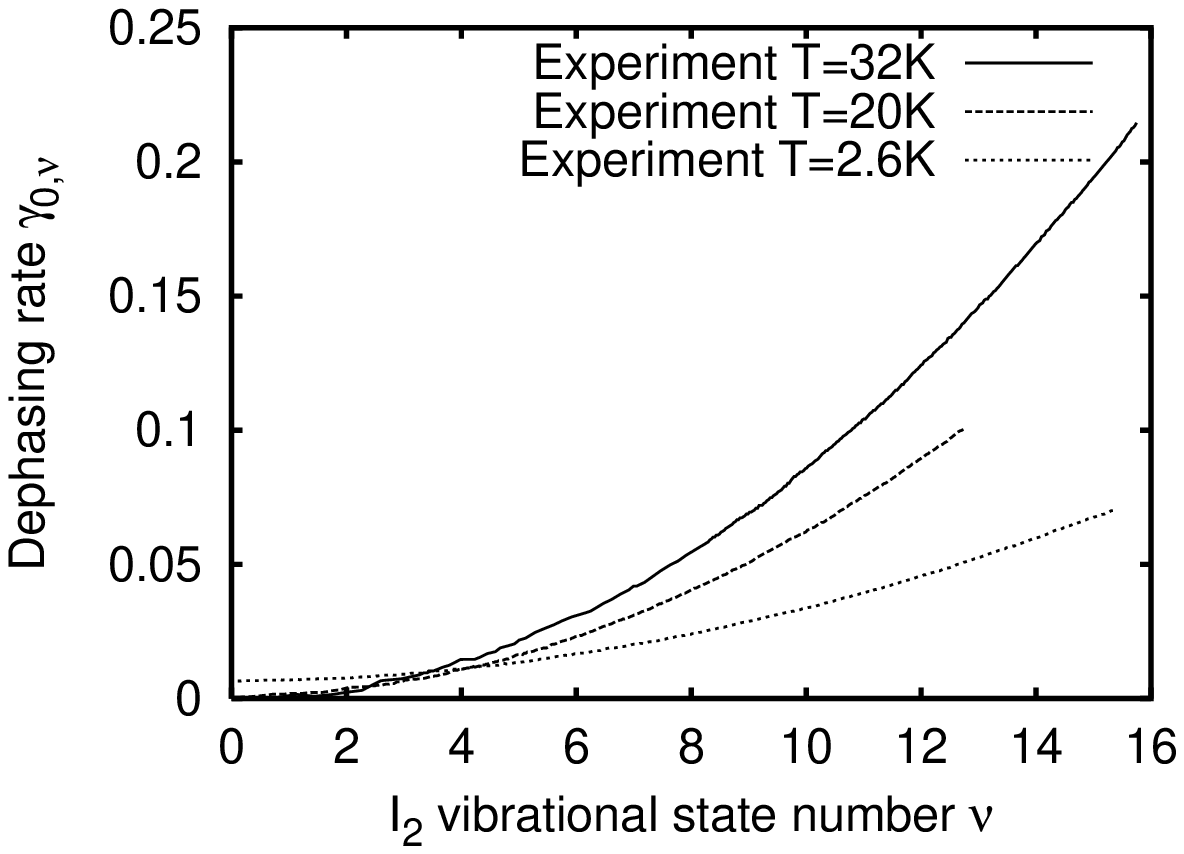} \includegraphics[width=3in]{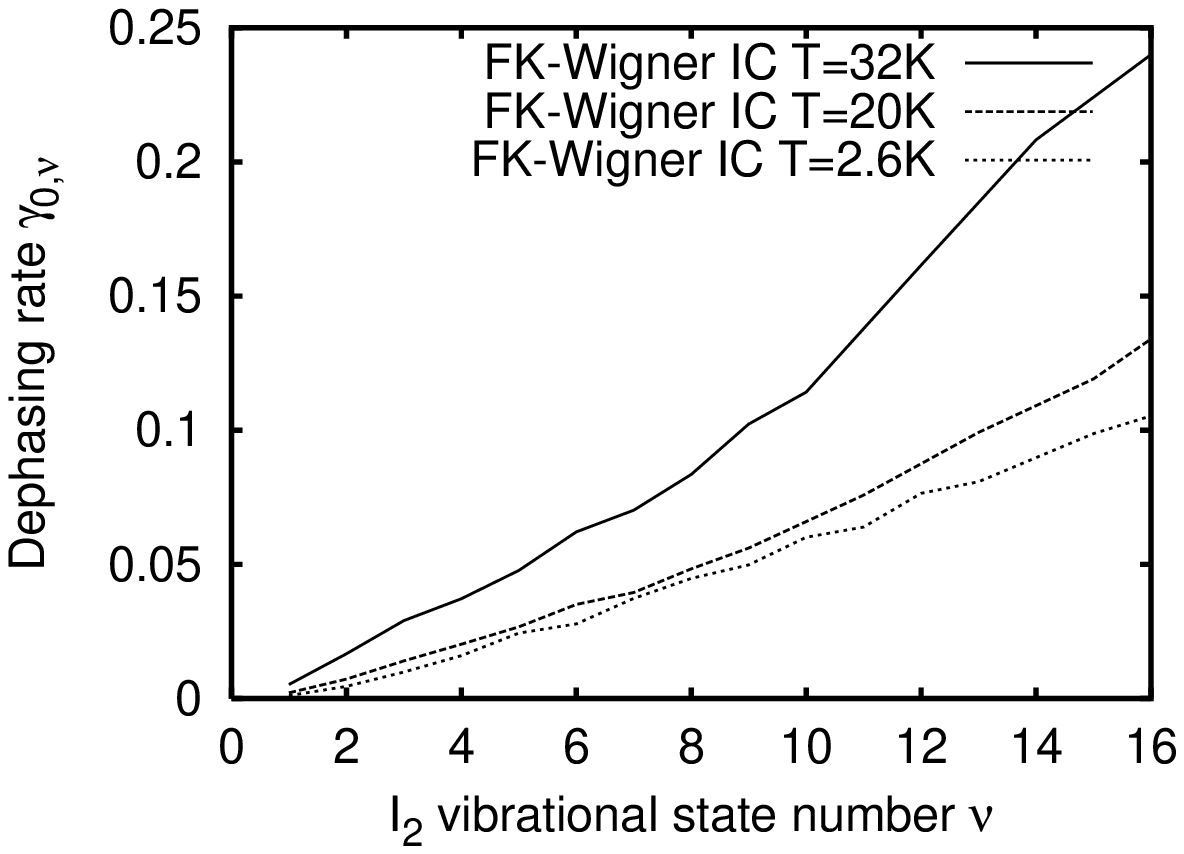}}
\caption{\label{fig:exp-Q-gammas}Comparison of experimental (left panel) and calculated (right panel) vibrational dephasing rates as functions of initial vibrational state superposition $(0,\nu)$ for iodine in solid krypton.}
\end{figure*}

In Fig. \ref{fig:grQ} we see that the quantum structure is fairly insensitive to a more than ten fold increase in temperature (T=2.6K to T=32K). The width of the pair distribution function peaks in this figure increase monotonically by about 20\% over this temperature range. Figure \ref{fig:exp-Q-gammas} compares experimental and calculated dephasing rates which show comparable monotonic increases with increasing temperature over this range. 

\section{\label{sec:conclusion}Conclusion} 
Approximate quantum dynamics methods which are based on combining path integral expressions for forward and backward propagators and linearizing in the difference between these paths provide a popular approach for incorporating some quantum effects with in a trajectory based framework. In this paper this type of approach has been implemented to study evolution of the off-diagonal elements of the vibrational density matrix which describe vibrational pure dephasing. In this approach trajectories of ``bath'' variables that influence the quantum subsystem evolve on the ``mean potential'' associated with the relevant states. This quasi-classical mean potential dynamics emerges naturally from the linearization theory which also gives a unique prescription for  the initial conditions of these trajectories. With this approach the Wigner transform of the full quantum equilibrium density operator is found to provide the initial phase space distribution for the quasi-classical trajectories. In this paper we have explored the accuracy of various methods for obtaining approximations to this quantum initial density. Our measure of quality here is the reliability with which these different methods can reproduce vibrational pure dephasing dynamics over a range of temperatures. Thus, for example, we found that the classical initial phase space distribution gives pure dephasing rates that agree with experiments in higher temperature crystals. As a consistency check the FK-Wigner approach for approximately sampling the full quantum phase space distribution which is developed in detail in this paper gives results that also reproduce the experimental findings in these higher temperature solids. However, in low temperature solids we find that the classical initial phase space distribution combined with the quasi-classical dynamics can not reproduce the experimental dephasing dynamics. When the quasi-classical trajectory initial conditions, however, are sampled from the FK-Wigner approximate quantum equilibrium density excellent agreement between experimental and calculated dephasing rates is found. 

There has been considerable theoretical work exploring vibrational dephasing in condensed phases. For example, Zewail and co-workers give a detailed exposition of the perturbation theory results applied to the anharmonic oscillator in a harmonic bath model \cite{Zewail:VibDe1,Zewail:VibDe2,Zewail:Relax}. Skinner and co-workers explored dephasing for a model two level system coupled to a harmonic bath where they obtained non-perturbative results \cite{Skinner01,Skinner02,Skinner04}. Generally when the Debye-like spectral density is employed in these different theoretical results the dephasing rates they predict have a very strong temperature dependence with $\gamma \sim T^7$. Similar sensitivity of dephasing rates to spectral density has been found in numerical simulations employing the stochastic classical trajectory approach \cite{Nitzan78, Nitzan78a} which ignores quantum effects. As discussed in section \ref{sec:Results} our realistic microscopic model calculations, which make no assumptions about interaction strength, anharmonicity, or assumptions about the nature of the underlying spectral density, and include quantum nuclear initial conditions effects, reproduce the very weak temperature dependence observed in the experiments. Increasing the temperature by about a factor of 10 (2.6K to 32K) results in a modest increase in dephasing by about a factor of two for most vibrational levels. Using their non-perturbative theory and a pseudo-local mode spectral density model which assumes that the vibrational dephasing occurs by coupling to low frequency modes that arise from, for example, hindered rotation and translation of the vibrating impurity in the crystal, Skinner and co-workers (and others \cite{Harris77,Harris79,Harris80,Wiersma79}) obtained various Arrhenius like forms for the temperature dependence of the dephasing rate. Such forms are consistent with the much weaker temperature dependence that we find in our calculations and in the experiments. A ``back of the envelope'' estimate using a single Arrhenius form, and assuming a temperature independent pre-exponential factor gives an activation energy or pseudo-local mode frequency of about 20 cm$^{-1}$. In future work we will explore the nature of these pseudo-local modes responsible for this dephasing dynamics by examining the microscopic motions responsible for dephasing in our realistic simulation model \cite{Ma08}.  

The study of dephasing dynamics in condensed phase systems is just one example that highlights the importance of reliable methods for including quantum effects in condensed phase calculations. With in the framework of linearization these quantum effects are incorporated principally in the initial distribution sampling. At present there are no direct ways to generate the Wigner transform of the thermal density operator and sample the exact full quantum distribution. Thus all current methods for implementing linearized dynamics for general applications use local Harmonic approximations. As shown in section \ref{sec:IC-sampling} the FK-Wigner approach employs a thermodynamic variational calculation to parameterize a locally quadratic approximate form for the Euclidean action appearing in the partition function. This form is then employed to approximate the Wigner transform of the density operator. With such an approach no local harmonic approximation is made to the potential. Thus the method incorporates global information from the partition function integral in order to fit the gaussian approximate form for the density. This leads to a form for the density operator that contains frequencies computed from a gaussian smearing of the interactions which can be implemented in a very computationally efficient means.

This situation is contrasted with that of alternative methods such as that of Shi and Geva \cite{ShiGeva03a}, in which a local Harmonic approximation to the bare potential is made. While these methods can use full PIMC calculations to accurately represent diagonal elements of the density matrix, using a local harmonic approximation for the off-diagonal elements can degrade the results. Thus the studies outlined in this paper suggest that the FK-Wigner approach is the method of choice for a balance of computational efficiency and accuracy.

\begin{acknowledgments}
This work was partially supported by a grant from the National Science Foundation (CHE-0616952). DFC acknowledges a travel grant from HPC-Europa and the hospitality of Prof. Ciccotti during his collaborative visit to the University of Rome, ``La Sapienza''.
We also acknowledge a grant of supercomputer time from Boston University's office of information technology and scientific computing and visualization. 
\end{acknowledgments}

\appendix
\section{\label{app:avgpots}Computation of potential averages over trial variational distribution}

In order to apply the variational result in Eq.(\ref{eq:var}) the various potentials averaged over the trial density defined in Eqs.(\ref{eq:Sdiff}) - (\ref{eq:avint}) must be computed. These calculations are most conveniently performed in Fourier space, thus we write
\begin{equation}
V(R(0)) = V(R) = \int_{-\infty}^{\infty} {dk \over (2 \pi)^d} {\tilde V}(k) e^{ik^TR}
\end{equation}
Further since $R(\tau) = {\bf M}^{-1/2} {\bf U} \eta(\tau)$ we find that 
\begin{widetext}
\begin{eqnarray}
\langle V(R) \rangle_{trial} & = &{1 \over Z_{trial}} \int {d\eta_c \over (2 \pi \hbar^2 \beta)^{d/2}} e^{-\beta L(\eta_c)} \prod_{n=1}^{\infty} \prod_{j=1}^d { \int d {\rm Re} \eta_{nj}  \int d {\rm Im} \eta_{nj} \over (\pi / \beta \Omega_n^2)^d}  \nonumber \\
\ & \ & \times e^{-\beta [\Omega_n^2 + \omega_j^2(\eta_c)] [({\rm Re} \eta_{nj})^2 + ({\rm Im} \eta_{nj})^2]}  \int_{-\infty}^{\infty} {dk \over (2 \pi)^d} {\tilde V}(k) e^{ik^TR_c} e^{2 \sum_j b_j({\rm Re}\eta_{nj})} 
\end{eqnarray}
\end{widetext}
with $b_j = i(k^T{\bf M}^{-1/2} {\bf U})_j$. The integrals over the real and imaginary parts of the $\eta_{nj}$ can be performed analytically using the same result that lead to the expression in Eq.(\ref{eq:ztrial}), and the final result has the form
\begin{eqnarray}
\langle V(R) \rangle_{trial} & = &{1 \over Z_{trial}} \int {dR_c \over (2 \pi \hbar^2 \beta)^{d/2}} e^{-\beta L(R_c)}\nonumber \\
\ & \ & \times  \prod_{j=1}^d {\beta \hbar \omega_j(R_c)/2 \over \sinh \beta \hbar \omega_j(R_c)/2} \nonumber \\
\ & \ & \times \int_{-\infty}^{\infty} {dk \over (2 \pi)^d} {\tilde V}(k) e^{ik^TR_c} e^{-{1 \over 2} k^T {\bf A}(R_c) k} \nonumber\\
\end{eqnarray}
where we have used the following 
\begin{eqnarray}
\sum_{j=1}^d \sum_{n=1}^{\infty} {(k^T {\bf M}^{-1/2} {\bf U})_j^2 \over \beta [\Omega_n^2 + \omega_j^2(R_c)] }& = &{1 \over 2} k^T {\bf M}^{-1/2} {\bf U} {\bf \Lambda} {\bf U}^T {\bf M}^{-1/2} k \nonumber \\
\ & \ & \times = {1 \over 2} k^T {\bf A} k
\end{eqnarray}
defining the smearing width matrix ${\bf A}$ as 
\begin{equation}
{\bf A} = {\bf M}^{-1/2} {\bf U} {\bf \Lambda} {\bf U}^T {\bf M}^{-1/2} \label{eq:A-def}
\end{equation}
with 
\begin{equation}
({\bf \Lambda})_{ij} = \sum_{n=1}^{\infty} {2 \over \beta[\Omega_n^2 + \omega_j^2(R_c)] } \delta_{ij} \label{eq:Lambda-def}
\end{equation}
Since $\tilde{V}(k) = FT\{V\}$ and $e^{-{1 \over 2} k^T {\bf A}(R_c) k} = FT\{|2\pi {\bf A}|^{-1/2} \exp[-{1 \over 2} R^T {\bf A}^{-1} R]\}$, and defining $V_A(R_c)$ as follows, we see that 
\begin{eqnarray}
&&V_A(R_c) = \int_{-\infty}^{\infty} {dk \over (2 \pi)^d} {\tilde V}(k) e^{ik^TR_c} e^{-{1 \over 2} k^T {\bf A}(R_c) k} = \nonumber \\
&& FT^{-1}\left \{FT\{V\} * FT\{|2\pi {\bf A}|^{-1/2} \exp[-{1 \over 2} R^T {\bf A}^{-1} R]\}\right \}(R_c) \label{eq:VA-kspace}\nonumber \\
\end{eqnarray}
Using the Fourier convolution theorem ($FT\{ f_1*f_2\} = FT\{ f_1\} FT\{ f_2\}$, where $(f_1*f_2)(q) = \int dy f_1(y) f_2(q-y)$) we can write
\begin{eqnarray}
\langle V(R) \rangle_{trial}&=&{1 \over Z_{trial}} \int {dR_c \over (2 \pi \hbar^2 \beta)^{d/2}} e^{-\beta L(R_c)}\nonumber\\
&\times & \prod_{j=1}^d {\beta \hbar \omega_j(R_c)/2 \over \sinh \beta \hbar \omega_j(R_c)/2} V_A(R_c)
\end{eqnarray}
so $V_A(R_c)$, defined above, can be interpreted as the Gaussian smeared potential since 
\begin{equation}
V_A(R_c) = \int_{-\infty}^{\infty} {dR \over |2\pi {\bf A}|^{1/2}} e^{-{1 \over 2} (R-R_c)^T {\bf A}^{-1}(R_c)(R-R_c)} V(R) \label{eq:VA-rspace}
\end{equation}
We can compute the Gaussian smeared local harmonic approximate trial potential in a similar fashion obtaining 
\begin{eqnarray}
&&V_A^{trial}(R_c) =  \int_{-\infty}^{\infty} {dR \over |2\pi {\bf A}|^{1/2}} e^{-{1 \over 2} (R-R_c)^T {\bf A}^{-1}(R_c)(R-R_c)} \nonumber \\
&& \times \left [{1 \over 2}(R-R_c)^T {\bf M}^{1/2} K(R_c) {\bf M}^{1/2}(R-R_c) + L(R_c) \right] \nonumber \\
&& = {1 \over 2} \sum_{ij} M_i^{1/2} K_{ij} M_j^{1/2} A_{ij} + L(R_c) 
\end{eqnarray}

\section{\label{app:variational}Variational Calculation Details}

To simplify notation we define 
\begin{equation}
P_{\beta}(R_c) = {1 \over (2 \pi \hbar^2 \beta)^{d/2}} \prod_{j=1}^d {\beta \hbar \omega_j(R_c)/2 \over \sinh \beta \hbar \omega_j(R_c)/2}
\end{equation}
and 
\begin{equation}
\chi(R_c)= {1 \over 2} \sum_{ij} M_i^{1/2} K_{ij} A_{ij} M_j^{1/2}
\end{equation}
so we can write 
\begin{equation}
Z_{trial}= \int dR_c e^{-\beta L(R_c)}P_{\beta}(R_c)
\end{equation}
Following Feynman and Kleinert \cite{Kleinert86} we proceed to optimize the right hand side of Eq.(\ref{eq:var}), $f=\exp[-{1 \over \hbar}\langle S - S_{trial} \rangle_{trial}] Z_{trial} = e^{-I/Z_{trial}} Z_{trial}$, with respect to the variational parameter functions ${\bf K}(R_c)$ and $L(R_c)$. Here we have defined 
\begin{equation}
I= \beta \int dR_c [V_A(R_c) - \chi(R_c) - L(R_c)] e^{-\beta L(R_c)}P_{\beta}(R_c)
\end{equation}
Partial functional differentiation of $f[{\bf K},L]$ first with respect to $L$ (fixed ${\bf K}$) yields
\begin{equation}
\delta _L f = e^{-I/Z_{trial}} [(I/Z_{trial} +1) \delta_L Z_{trial} - \delta_L I]
\end{equation}
and since, for example, 
\begin{equation}
\delta_L I = \int dR_c {\delta I \over \delta L(R_c)}(R_c) \delta L(R_c)
\end{equation}
Where the partial derivative notation means regard the functional $I[L(R_c)]=I(L(R_c^1),L(R_c^2),\dots, L(R_c^n))$ in the limit $n\rightarrow \infty$ as a multidimensional function of $L$ evaluated at various points $R_c^i$, and differentiate with respect to $L(R_c)$ at the point $R_c$ while keeping all other values of $L(R_c'\ne R_c)$, and the values of the functions ${\bf K}$ at all points fixed. Thus we find 
\begin{eqnarray}
\delta_L f & = & e^{-I/Z_{trial}} \{\beta^2 \int dR_c  [V_A(R_c) - \chi(R_c) - L(R_c)]\nonumber \\
\ & \ &  e^{-\beta L(R_c)}P_{\beta}(R_c) \delta L(R_c)\nonumber \\
\ & \ & - \beta I/Z_{trial} \int dR_c  e^{-\beta L(R_c)}P_{\beta}(R_c) \delta L(R_c) \}
\end{eqnarray}
The only way to get both terms in the curly brackets in this expression for $\delta_L f$ to vanish for arbitrary $\delta L(R_c)$ is if the quantity $[V_A(R_c) - \chi(R_c) - L(R_c)]$ is zero for each $R_c$, giving the following local  relationship between the parameter functions characterizing the extrema with respect to $L$
\begin{equation}  
L_e(R_c) = V_A(R_c) - {1 \over 2} \sum_{ij} M_i^{1/2} K_{ij}(R_c) A_{ij}(R_c) M_j^{1/2} \label{eq:varL}
\end{equation}
With this result $I=0$, and we now must optimize $f[{\bf K},L_e]=Z_{trial}[{\bf K},L_e]$ with respect to variations in the functions ${\bf K}$ with $L$ fixed at $L_e$. Now
\begin{equation}
Z_{trial}[{\bf K}]=\int {dR_c \over (2 \pi \hbar^2 \beta)^{d/2}} e^{-\beta W_{trial}(R_c)}
\end{equation}
and $W_{trial}$ for $L=L_e$ is obtained by substituting Eq.(\ref{eq:varL}) into Eq.(\ref{eq:Wtrial}) to give the following expression for the trial effective potential that depends only on the local curvature parameter matrix ${\bf K}$
\begin{eqnarray}
W_{trial}(R_c) & = & -{1 \over \beta} \sum_{j=1}^d \ln \left({\beta\hbar\omega_j(R_c)/2 \over \sinh \beta\hbar\omega_j(R_c)/2} \right) \label{eq:Wtrial-fin} \\
\ & \ & + V_A(R_c) - {1 \over 2} \sum_{ij} M_i^{1/2} K_{ij}(R_c) A_{ij}(R_c) M_j^{1/2} \nonumber
\end{eqnarray}
Functional optimization with respect to each of the functions $K_{ij}$ requires
\begin{eqnarray}
\delta_{K_{ij}} Z_{trial}&=& \int dR_c {\delta Z_{trial} \over \delta K_{ij}(R_c)} \delta K_{ij}(R_c) \nonumber \\
\ & \ & = -\beta \int dR_c {\delta W_{trial} \over \delta K_{ij}(R_c)} e^{-\beta W_{trial}(R_c)} \delta K_{ij}(R_c)  \nonumber \\
\ & \ &= 0
\end{eqnarray}
For this to be true for arbitrary variation functions $\delta K_{ij}(R_c)$, $(\partial W_{trial} / \partial K_{ij})(R_c)$ must vanish locally for each function $K_{ij}$. Since ${\bf A}$ is a complicated function of ${\bf K}$ given by Eq.(\ref{eq:A-def}) we write $W_{trial} = W_{trial}({\bf K},{\bf A}({\bf K}))$ and apply the chain rule to obtain
\begin{equation}
{\partial W_{trial} \over \partial K_{ij}} = \left( \partial W_{trial} \over \partial K_{ij} \right)_{\bf A} + \left( \partial W_{trial} \over \partial A_{ij} \right)_{\bf K} {\partial A_{ij} \over \partial K_{ij}} \label{eq:dwdk}
\end{equation}
After some straightforward manipulations we find that  
\begin{eqnarray}
&&\left({\partial W_{trial} \over \partial K_{j'k'}} \right)_{\bf A} = -{1 \over 2} M_{j'}^{1/2} A_{j'k'} M_{k'}^{1/2} \nonumber \\
&&+ \sum_{l=1}^d{1 \over \beta \omega_l}\{ {\beta \hbar \omega_l \over 2} \coth {\beta \hbar \omega_l \over 2} -1 \} \left({\partial \omega_l \over \partial K_{j'k'}} \right)_{\bf A}
\end{eqnarray}
Further since  $\omega^2 = {\bf U}^T {\bf K} {\bf U}$ is diagonal, so $\omega_l = [ \sum_j \sum_k U_{lj}^T K_{jk} U_{kl}]^{1/2}$ and thus 
\begin{equation}
\left({\partial \omega_l \over \partial K_{j'k'}} \right)_{\bf A} = {1 \over 2 \omega_l} U_{lj'}^T U_{k'l}
\end{equation}
Moreover, since 
\begin{equation}
\coth \pi x = {1 \over \pi x} + {2 x \over \pi} \sum_{n=1}^{\infty} {1 \over x^2 + n^2}
\end{equation}
it is easily shown that 
\begin{equation}
\Lambda_{ll} = \sum_{k=1}^{\infty} {2 \over \beta[\Omega_k^2+ \omega_l^2]} = {1\over \beta \omega_l^2} \{ {\beta \hbar \omega_l \over 2} \coth {\beta \hbar \omega_l \over 2} -1 \} \label{eq:Lambda-def-1}
\end{equation}
and thus the first term on the right hand side of Eq.(\ref{eq:dwdk}) vanishes identically using the results around Eq.(\ref{eq:A-def}). Thus Eq.(\ref{eq:dwdk}) gives the optimization condition as $(\partial W_{trial} / \partial A_{ij})_{\bf K} = 0$, since in general $\partial A_{ij} / \partial K_{ij}$ is nonzero. 

Differentiating $W_{trial}$ with respect to a particular element of ${\bf A}$ keeping all others fixed (and pretending that ${\bf K}$ can be held fixed during this process consistent with the application of the chain rule) we find, using the Fourier space form of $V_A$ in Eq.(\ref{eq:VA-kspace}), that
\begin{eqnarray}
{\partial W_{trial} \over \partial A_{j'k'}}& =& -{1\over 2} \{ \int {d{\bf k} \over (2\pi)^d} e^{i{\bf k}^TR_c}{\tilde V}({\bf k}) k_{j'} k_{k'} e^{-{1 \over 2} {\bf k}^T {\bf A}(R_c) {\bf k}} \nonumber \\
\ & \ & + M_{j'}^{1/2} K_{j'k'} M_{k'}^{1/2} \}
\end{eqnarray}
Packing all these terms into a $d \times d$ matrix, the extremum condition requires that all elements are zero and the result can be written in matrix form as follows
\begin{equation}
-\int {d{\bf k} \over (2\pi)^d} e^{i{\bf k}^TR_c}{\tilde V}({\bf k}) {\bf k} {\bf k}^T e^{-{1 \over 2} {\bf k}^T {\bf A}(R_c) {\bf k}} = {\bf M}^{1/2} {\bf K} {\bf M}^{1/2} \label{eq:var-fin}
\end{equation}
It is easy to show that
\begin{equation}
FT\left \{ {\partial^2 V \over \partial R \partial R^T} \right \} = {\tilde V}({\bf k}) (i{\bf k})(i{\bf k})^T
\end{equation}
So using the same manipulations based on the convolution theorem that lead to the result in Eq.(\ref{eq:VA-rspace}) we obtain
\begin{equation}
{\bf K}(R_c) = \int  {dR \over |2\pi {\bf A}|^{1/2}} {\bf D}(R) e^{-{1 \over 2} (R-R_c)^T {\bf A}^{-1}(R_c)(R-R_c)} 
\end{equation}
Where ${\bf D}(R)$ is the mass weighted Hessian 
\begin{equation}
{\bf D}(R) ={\bf M}^{-1/2} \left( {\partial^2 V \over \partial R \partial R^T}  \right)(R) {\bf M}^{-1/2}
\end{equation}

\section{\label{app:smear-pair}General approach for Gaussian smearing pair potentials}

The left hand side of Eq.(\ref{eq:var-fin}) is conveniently evaluated in Fourier space if the interaction potential is pairwise additive $V(R) = \sum_{i < j} v^{ij}(r_{ij})$, and each pair function of the inter-particle distance $r_{ij}=|R_i - R_j|$ is fit to a sum of Gaussians $v^{ij}(r_{ij}) = \sum_k a^{ij}_k \exp(-{1 \over 2} b^{ij}_k r_{ij}^2)$. In ${\bf k}$-space this potential has the form
\begin{eqnarray} 
{\tilde V}({\bf k})&=& \sum_{i < j} \sum_k (2 \pi)^{3(N-1)} \left( \prod_{l \ne i,j} \delta({\bf k}_l) \right) a^{ij}_k \left({2 \pi \over b^{ij}_k} \right)^{3/2}\nonumber\\
& &\times e^{-|{\bf k}_i|^2/2b^{ij}_k} \delta({\bf k}_i+{\bf k}_j)
\end{eqnarray}
Using this result we can rewrite the general $(mx,ny)$ matrix element of Eq.(\ref{eq:var-fin}) giving
\begin{widetext}
\begin{eqnarray}
(M_m M_n)^{1/2} K_{mx,ny} & = & \sum_k a^{mn}_k (2 \pi b^{mn}_k)^{-3/2} \int d{\bf k}_m e^{i{\bf k}_m^T (R_{cm}-R_{cn})} k_{mx} k_{my} e^{-|{\bf k}_i|^2/2b^{mn}_k} \nonumber \\
\ & \ & \times \exp[-{1 \over 2} \sum_{i=1}^{3} \sum_{j=1}^{3} k_{mi} ([A_{mimj} + A_{ninj}] - [A_{minj}+ A_{nimj}])k_{mj}] \nonumber 
\end{eqnarray}
\end{widetext}
Constructing the symmetric $3 \times 3$ matrix ${\bf \gamma}^{nm}$ for particles $n$ and $m$ with elements of the form
\begin{equation}
{\bf \gamma}_{xx}^{nm} = [A_{mxmx} + A_{nxnx}] - [A_{mxnx}+ A_{nxmx}] + 1/b^{mn}_k
\end{equation}
and 
\begin{equation}
{\bf \gamma}_{xy}^{nm} =[A_{mxmy} + A_{nxny}] - [A_{mxny}+ A_{nxmy}]
\end{equation}
the remaining ${\bf k}_m$ integral can be readily performed yielding the following expression for the $3 \times 3$ ${\bf K}^{mn}$ sub-matrix
\begin{widetext}
\begin{eqnarray}
{\bf K}^{mn} & = & (M_m M_n)^{-1/2}\sum_k {a^{mn}_k \over \sqrt{(b^{mn}_k)^3 \det \gamma^{nm}}} [(\gamma^{mn})^{-1}-{\bf u}^{mn} ({\bf u}^{mn})^T] \nonumber \\
\ & \ & \times \exp[-{1\over 2} (R_{cm}-R_{cn})^T (\gamma^{mn})^{-1} (R_{cm} - R_{cn})] \label{eq:K-fin}
\end{eqnarray}
\end{widetext}
Where ${\bf u}^{mn} = (\gamma^{mn})^{-1} (R_{cm}-R_{cn})$. This approach is readily simplified to compute $V_A(R_c)$. Proceeding in a similar fashion to that outlined above employing the ${\bf k}$-space expression of Eq.(\ref{eq:VA-kspace}) we obtain
\begin{eqnarray}
V_A(R_c)&=& \sum_{m<n} \sum_k {a^{mn}_k \over \sqrt{(b^{mn}_k)^3 \det \gamma^{nm}}} \nonumber\\
&\times &\exp[-{1\over 2} (R_{cm}-R_{cn})^T (\gamma^{mn})^{-1} (R_{cm} - R_{cn})]\nonumber\\
\end{eqnarray}

\section{\label{app:rhoH-WT}Local Harmonic Approximate Density Operator and its Wigner Transformation}

Thus we define the density operator with the variationally optimized trial harmonic form as
\begin{widetext}
\begin{eqnarray}
e_{trial}^{-\beta {\hat H}} & = & \int dq' \int dq'' |q' \rangle \langle q''|  \int dq_c \int_{q(0) = q'}^{q(\beta\hbar) = q''} {\cal D}[q(\tau)] \delta({1 \over \beta \hbar} \int_0^{\beta \hbar} q(\tau) d\tau-q_c) \nonumber \\ 
\ & \ & \times \exp[- {1\over \hbar} \int_0^{\beta \hbar} d\tau \{ {1\over 2}{\dot q}(\tau)^T {\dot q}(\tau) + L(q_c)+ {1\over 2} (q(\tau) - q_c)^T {\bf K}(q_c) (q(\tau) - q_c) \}]
\end{eqnarray}
\end{widetext}
or writing the $\delta$-function in integral representation we have
\begin{widetext}
\begin{eqnarray}
e_{trial}^{-\beta {\hat H}} & = & \int dq' \int dq'' |q' \rangle \langle q''| \int dq_c \int {dk \over (2 \pi)^d} \int_{q(0) = q'}^{q(\beta\hbar) = q''} {\cal D}[q(\tau)] \\
& \ & \times \exp[-{1 \over \hbar} \int_0^{\beta \hbar} d\tau \{ {1\over 2}{\dot q}(\tau)^T {\dot q}(\tau) + L(q_c)+ {1\over 2} (q(\tau) - q_c)^T {\bf K}(q_c) (q(\tau) - q_c) + ({i \over \beta}) k^T (q(\tau) - q_c) \}] \nonumber
\end{eqnarray}
\end{widetext}
Transforming to normal mode vectors $\eta={\bf U}^T q$ where ${\bf U}$ diagonalizes the variationally optimized local curvature matrix so that ${\bf K}(q_c) = {\bf U} \omega^2(q_c) {\bf U}^T$ as in Eq.(\ref{eq:Z-trial-normal}), and defining $y(\tau) = (\eta(\tau)-\eta_c)$ this expression becomes
\begin{widetext}
\begin{eqnarray}
e_{trial}^{-\beta {\hat H}} & = & \int d\eta' \int d\eta'' |\eta' \rangle \langle \eta''| \int d\eta_c 
e^{-\beta L(\eta_c)}\int {d\kappa \over (2 \pi)^d} \int_{y(0) = (\eta'-\eta_c)}^{y(\beta\hbar) = (\eta''-\eta_c)} {\cal D}[y(\tau)] \nonumber \\
\ & \ & \times \exp[-{1 \over \hbar} \int_0^{\beta \hbar} d\tau {1\over 2}\{ {\dot y}(\tau)^T {\dot y}(\tau)
+y(\tau)^T {\bf \omega}^2 y(\tau) + ({2 i \over \beta}) \kappa^T y(\tau) \}] \label{eq:Boltz}
\end{eqnarray}
where $\kappa= {\bf U}^T k$. However, the path integral  
\begin{eqnarray}
I = \int_{x(0)=x}^{x(\beta \hbar) =x'} {\cal D}[x(\tau)] \exp[-{1 \over \hbar} \int_0^{\beta \hbar} d\tau {1\over 2}\{ {\dot x}^2(\tau)
+\omega^2 x^2(\tau) + 2 a x(\tau) \}] \nonumber
\end{eqnarray}
can be readily evaluated yeilding
\begin{eqnarray}
I & = & \left(\omega \over 2 \pi \hbar \sinh(\beta \hbar \omega) \right)^{1/2} \exp[-{\omega (x'^2 + x^2) \over 2 \hbar \tanh(\beta \hbar \omega)} + {\omega x x' \over \hbar \sinh(\beta \hbar \omega)}] \\
\ & \ & \times \exp[{a^2 \over \hbar \omega^3} \{(\beta \hbar \omega /2) - \tanh(\beta \hbar \omega /2)\} + {a \over \hbar \omega} (x+x') \tanh(\beta \hbar \omega/2)] \nonumber
\end{eqnarray}
\end{widetext}
With $a= 2 i \kappa_n / \beta$ as in Eq.(\ref{eq:Boltz}), $I$ has a Gaussian form in $\kappa_n$ and the Gaussian integrals over these components in Eq.(\ref{eq:Boltz}) can be performed analytically. The final result of these manipulations is 
\begin{widetext}
\begin{eqnarray}
e_{trial}^{-\beta {\hat H}} & = & \int d\eta' \int d\eta'' \int {d\eta_c \over (2 \pi)^d}
e^{-\beta W_{trial}(\eta_c)} |\eta' \rangle \langle \eta''| \nonumber \\
\ & \ & \times \prod_i [ \exp \left\{ -{\omega_i \over \hbar \alpha_i} ({\eta_i'' + \eta_i' \over 2} - \eta_{ci})^2 - {\omega_i \over 4 \hbar} \coth(\beta \hbar \omega_i / 2) (\eta_i'' - \eta_i')^2 \right\}
{\sqrt{\omega_i \over \pi \hbar \alpha_i}} {\sqrt{2 \pi \over \hbar^2 \beta}}]
\label{eq:Boltz-fin}
\end{eqnarray}
\end{widetext}
Here we define the quantity 
\begin{equation}
\alpha_i = \coth \beta \hbar \omega_i/2 - 2/\beta \hbar \omega_i
\end{equation}

The above locally Gaussian form in the normal mode variables $\eta'$, and $\eta''$ for the Boltzmann operator is particularly convenient for approximately evaluating thermally averaged time correlation functions in the linearized approximation \cite{Coker05a,Coker05b,Poulsen03} where quantities like Wigner transforms of products of operators of interest with the Boltzmann operator {\em i.e.} 
\begin{equation}
({\hat \rho}_{\beta} {\hat O})_W(Q,P) = \int dz \langle Q + z/2 | {\hat \rho}_{\beta} {\hat O} | Q - z/2 \rangle e^{-{i\over \hbar} Pz}
\end{equation} 
are required. For simplicity we will focus on the Wigner transform of the Boltzmann operator itself. In the applications to computing linearized approximations to time correlation functions mentioned above the quantity $({\hat \rho}_{\beta} {\hat O})_W(Q,P)$ serves both to provide a distribution for initial conditions for classical-like trajectories and includes the measurement of the initial operator. Depending on the nature of the initial operator ${\hat O}$, the product Wigner transform may be dominated by the Wigner transform of the Boltzmann operator. This is the case in the applications to vibrational dephasing that we will report in a subsequent publication \cite{Coker08} as the operator in question just involves the vibrational states of the solute chromophore. The solvent initial degrees of freedom must thus be sampled from the Wigner transform of the solvent Boltzmann operator. Here we thus explore the accuracy of the Feynman-Kleinert approximation for the initial Wigner phase space distribution and the underlying approximation to the thermal density matrix. The Wigner transform of the trial Boltzmann operator written in the local normal mode phase space representation thus has the form
\begin{widetext}
\begin{eqnarray}
(e_{trial}^{-\beta {\hat H}})_W(Q,P_Q) & = & \int {d\eta_c \over (2 \pi \hbar)^d} e^{-\beta W_{trial}(\eta_c)}
\prod_{i=1}^d \{ \left( {4\pi \over \coth(\beta \hbar \omega_i/2) \alpha_i \beta/2} \right)^{1/2} \label{eq:Boltz-Wigner}\\
\ & \ & \times \exp[-{\omega_i(\eta_c)\over \alpha_i \hbar} (Q_i-\eta_{ci})^2] \exp[-{\tanh(\beta\hbar\omega_i/2) \over \hbar \omega_i} P_{Qi}^2] \} \nonumber
\end{eqnarray} 
 \end{widetext}

\bibliography{FK-ic-sampling}

\begin{thebibliography}{76}
\expandafter\ifx\csname natexlab\endcsname\relax\def\natexlab#1{#1}\fi
\expandafter\ifx\csname bibnamefont\endcsname\relax
  \def\bibnamefont#1{#1}\fi
\expandafter\ifx\csname bibfnamefont\endcsname\relax
  \def\bibfnamefont#1{#1}\fi
\expandafter\ifx\csname citenamefont\endcsname\relax
  \def\citenamefont#1{#1}\fi
\expandafter\ifx\csname url\endcsname\relax
  \def\url#1{\texttt{#1}}\fi
\expandafter\ifx\csname urlprefix\endcsname\relax\def\urlprefix{URL }\fi
\providecommand{\bibinfo}[2]{#2}
\providecommand{\eprint}[2][]{\url{#2}}

\bibitem[{\citenamefont{Diestler and Zewail}({1979a})}]{Zewail:VibDe1}
\bibinfo{author}{\bibfnamefont{D.~J.} \bibnamefont{Diestler}} \bibnamefont{and}
  \bibinfo{author}{\bibfnamefont{A.~H.} \bibnamefont{Zewail}},
  \bibinfo{journal}{J. Chem. Phys.} \textbf{\bibinfo{volume}{{\bf 71}}},
  \bibinfo{pages}{3103} (\bibinfo{year}{{1979a}}).

\bibitem[{\citenamefont{Diestler and Zewail}({1979b})}]{Zewail:VibDe2}
\bibinfo{author}{\bibfnamefont{D.~J.} \bibnamefont{Diestler}} \bibnamefont{and}
  \bibinfo{author}{\bibfnamefont{A.~H.} \bibnamefont{Zewail}},
  \bibinfo{journal}{J. Chem. Phys.} \textbf{\bibinfo{volume}{{\bf 71}}},
  \bibinfo{pages}{3113} (\bibinfo{year}{{1979b}}).

\bibitem[{\citenamefont{Zewail and Diestler}({1979c})}]{Zewail:Relax}
\bibinfo{author}{\bibfnamefont{A.~H.} \bibnamefont{Zewail}} \bibnamefont{and}
  \bibinfo{author}{\bibfnamefont{D.~J.} \bibnamefont{Diestler}},
  \bibinfo{journal}{Chem. Phys. Lett.} \textbf{\bibinfo{volume}{{\bf 65}}},
  \bibinfo{pages}{37} (\bibinfo{year}{{1979c}}).

\bibitem[{\citenamefont{Skinner}({1988})}]{Skinner02}
\bibinfo{author}{\bibfnamefont{J.~L.} \bibnamefont{Skinner}},
  \bibinfo{journal}{Ann. Rev. Phys. Chem.} \textbf{\bibinfo{volume}{{\bf 39}}},
  \bibinfo{pages}{463} (\bibinfo{year}{{1988}}).

\bibitem[{\citenamefont{Skinner and Hsu}({1986})}]{Skinner01}
\bibinfo{author}{\bibfnamefont{J.~L.} \bibnamefont{Skinner}} \bibnamefont{and}
  \bibinfo{author}{\bibfnamefont{D.}~\bibnamefont{Hsu}}, \bibinfo{journal}{J.
  Phys. Chem.} \textbf{\bibinfo{volume}{{\bf 90}}}, \bibinfo{pages}{4931}
  (\bibinfo{year}{{1986}}).

\bibitem[{\citenamefont{Hsu and Skinner}({1985})}]{Skinner04}
\bibinfo{author}{\bibfnamefont{D.}~\bibnamefont{Hsu}} \bibnamefont{and}
  \bibinfo{author}{\bibfnamefont{J.~L.} \bibnamefont{Skinner}},
  \bibinfo{journal}{J. Chem. Phys.} \textbf{\bibinfo{volume}{{\bf 83}}},
  \bibinfo{pages}{2097} (\bibinfo{year}{{1985}}).

\bibitem[{\citenamefont{Weiss}({1999})}]{Weiss}
\bibinfo{author}{\bibfnamefont{U.}~\bibnamefont{Weiss}},
  \emph{\bibinfo{title}{Quantum Dissipative Systems}}
  (\bibinfo{publisher}{World Scientific}, \bibinfo{address}{Singapore},
  \bibinfo{year}{{1999}}).

\bibitem[{\citenamefont{Makarov and Metiu}({1999})}]{Makarov99}
\bibinfo{author}{\bibfnamefont{D.~E.} \bibnamefont{Makarov}} \bibnamefont{and}
  \bibinfo{author}{\bibfnamefont{H.}~\bibnamefont{Metiu}}, \bibinfo{journal}{J.
  Chem. Phys.} \textbf{\bibinfo{volume}{{\bf 111}}}, \bibinfo{pages}{10126}
  (\bibinfo{year}{{1999}}).

\bibitem[{\citenamefont{Dalibard et~al.}({1992})\citenamefont{Dalibard, Castin,
  and Molmer}}]{Dalibard92}
\bibinfo{author}{\bibfnamefont{J.}~\bibnamefont{Dalibard}},
  \bibinfo{author}{\bibfnamefont{Y.}~\bibnamefont{Castin}}, \bibnamefont{and}
  \bibinfo{author}{\bibfnamefont{K.}~\bibnamefont{Molmer}},
  \bibinfo{journal}{Phys. Rev. Letts.} \textbf{\bibinfo{volume}{{\bf 68}}},
  \bibinfo{pages}{580} (\bibinfo{year}{{1992}}).

\bibitem[{\citenamefont{Dum et~al.}({1992})\citenamefont{Dum, Zoller, and
  Ritsch}}]{Dum92}
\bibinfo{author}{\bibfnamefont{R.}~\bibnamefont{Dum}},
  \bibinfo{author}{\bibfnamefont{P.}~\bibnamefont{Zoller}}, \bibnamefont{and}
  \bibinfo{author}{\bibfnamefont{H.}~\bibnamefont{Ritsch}},
  \bibinfo{journal}{Phys. Rev. A} \textbf{\bibinfo{volume}{{\bf 45}}},
  \bibinfo{pages}{4879} (\bibinfo{year}{{1992}}).

\bibitem[{\citenamefont{Fredj et~al.}({1996})\citenamefont{Fredj, Gerber, and
  Ratner}}]{Gerber96}
\bibinfo{author}{\bibfnamefont{E.}~\bibnamefont{Fredj}},
  \bibinfo{author}{\bibfnamefont{R.}~\bibnamefont{Gerber}}, \bibnamefont{and}
  \bibinfo{author}{\bibfnamefont{M.}~\bibnamefont{Ratner}},
  \bibinfo{journal}{J. Chem. Phys.} \textbf{\bibinfo{volume}{{\bf 105}}},
  \bibinfo{pages}{1121} (\bibinfo{year}{{1996}}).

\bibitem[{\citenamefont{Jungwirth et~al.}({1997})\citenamefont{Jungwirth,
  Fredj, and Gerber}}]{Gerber97}
\bibinfo{author}{\bibfnamefont{P.}~\bibnamefont{Jungwirth}},
  \bibinfo{author}{\bibfnamefont{E.}~\bibnamefont{Fredj}}, \bibnamefont{and}
  \bibinfo{author}{\bibfnamefont{R.}~\bibnamefont{Gerber}},
  \bibinfo{journal}{J. Chem. Phys.} \textbf{\bibinfo{volume}{{\bf 107}}},
  \bibinfo{pages}{8963} (\bibinfo{year}{{1997}}).

\bibitem[{\citenamefont{Jungwirth and Gerber}({1999})}]{Gerber99}
\bibinfo{author}{\bibfnamefont{P.}~\bibnamefont{Jungwirth}} \bibnamefont{and}
  \bibinfo{author}{\bibfnamefont{R.}~\bibnamefont{Gerber}},
  \bibinfo{journal}{Chem. Rev. (Washington D.C.)} \textbf{\bibinfo{volume}{{\bf
  99}}}, \bibinfo{pages}{1583} (\bibinfo{year}{{1999}}).

\bibitem[{\citenamefont{Poulsen et~al.}({2003})\citenamefont{Poulsen, Nyman,
  and Rossky}}]{Poulsen03}
\bibinfo{author}{\bibfnamefont{J.~A.} \bibnamefont{Poulsen}},
  \bibinfo{author}{\bibfnamefont{G.}~\bibnamefont{Nyman}}, \bibnamefont{and}
  \bibinfo{author}{\bibfnamefont{P.~J.} \bibnamefont{Rossky}},
  \bibinfo{journal}{J. Chem. Phys.} \textbf{\bibinfo{volume}{{\bf 119}}},
  \bibinfo{pages}{12179} (\bibinfo{year}{{2003}}).

\bibitem[{\citenamefont{Shi and Geva}({2003b})}]{ShiGeva03b}
\bibinfo{author}{\bibfnamefont{Q.}~\bibnamefont{Shi}} \bibnamefont{and}
  \bibinfo{author}{\bibfnamefont{E.}~\bibnamefont{Geva}}, \bibinfo{journal}{J.
  Chem. Phys.} \textbf{\bibinfo{volume}{{\bf 118}}}, \bibinfo{pages}{8173}
  (\bibinfo{year}{{2003b}}).

\bibitem[{\citenamefont{Shi and Geva}({2003c})}]{ShiGeva03c}
\bibinfo{author}{\bibfnamefont{Q.}~\bibnamefont{Shi}} \bibnamefont{and}
  \bibinfo{author}{\bibfnamefont{E.}~\bibnamefont{Geva}}, \bibinfo{journal}{J.
  Chem. Phys.} \textbf{\bibinfo{volume}{{\bf 119}}}, \bibinfo{pages}{9030}
  (\bibinfo{year}{{2003c}}).

\bibitem[{\citenamefont{Shi and Geva}({2003d})}]{ShiGeva03d}
\bibinfo{author}{\bibfnamefont{Q.}~\bibnamefont{Shi}} \bibnamefont{and}
  \bibinfo{author}{\bibfnamefont{E.}~\bibnamefont{Geva}}, \bibinfo{journal}{J.
  Chem. Phys.} \textbf{\bibinfo{volume}{{\bf 118}}}, \bibinfo{pages}{7562}
  (\bibinfo{year}{{2003d}}).

\bibitem[{\citenamefont{Hernandez and Voth}({1998})}]{Hernandez98}
\bibinfo{author}{\bibfnamefont{R.}~\bibnamefont{Hernandez}} \bibnamefont{and}
  \bibinfo{author}{\bibfnamefont{G.~A.} \bibnamefont{Voth}},
  \bibinfo{journal}{Chem. Phys.} \textbf{\bibinfo{volume}{{\bf 233}}},
  \bibinfo{pages}{243} (\bibinfo{year}{{1998}}).

\bibitem[{\citenamefont{Miller}({2001})}]{Miller01-IVR}
\bibinfo{author}{\bibfnamefont{W.}~\bibnamefont{Miller}}, \bibinfo{journal}{J.
  Phys. Chem.} \textbf{\bibinfo{volume}{{\bf 105}}}, \bibinfo{pages}{2942}
  (\bibinfo{year}{{2001}}).

\bibitem[{\citenamefont{Causo et~al.}({2005b})\citenamefont{Causo, Ciccotti,
  Bonella, Montemayor, and Coker}}]{Coker05b}
\bibinfo{author}{\bibfnamefont{S.}~\bibnamefont{Causo}},
  \bibinfo{author}{\bibfnamefont{G.}~\bibnamefont{Ciccotti}},
  \bibinfo{author}{\bibfnamefont{S.}~\bibnamefont{Bonella}},
  \bibinfo{author}{\bibfnamefont{D.}~\bibnamefont{Montemayor}},
  \bibnamefont{and} \bibinfo{author}{\bibfnamefont{D.~F.} \bibnamefont{Coker}},
  \bibinfo{journal}{J. Phys. Chem. B} \textbf{\bibinfo{volume}{{\bf 109}}},
  \bibinfo{pages}{6855} (\bibinfo{year}{{2005b}}).

\bibitem[{\citenamefont{Martens and Fang}({1997})}]{Martens01}
\bibinfo{author}{\bibfnamefont{C.~C.} \bibnamefont{Martens}} \bibnamefont{and}
  \bibinfo{author}{\bibfnamefont{J.-Y.} \bibnamefont{Fang}},
  \bibinfo{journal}{J. Chem. Phys.} \textbf{\bibinfo{volume}{{\bf 106}}},
  \bibinfo{pages}{4918} (\bibinfo{year}{{1997}}).

\bibitem[{\citenamefont{Almy et~al.}({2000})\citenamefont{Almy, Kizer, Zadoyan,
  and Apkarian}}]{Apkarian00}
\bibinfo{author}{\bibfnamefont{J.}~\bibnamefont{Almy}},
  \bibinfo{author}{\bibfnamefont{K.}~\bibnamefont{Kizer}},
  \bibinfo{author}{\bibfnamefont{R.}~\bibnamefont{Zadoyan}}, \bibnamefont{and}
  \bibinfo{author}{\bibfnamefont{V.~A.} \bibnamefont{Apkarian}},
  \bibinfo{journal}{J. Phys. Chem. A} \textbf{\bibinfo{volume}{{\bf 104}}},
  \bibinfo{pages}{3508} (\bibinfo{year}{{2000}}).

\bibitem[{\citenamefont{Karavitis et~al.}({2001a})\citenamefont{Karavitis,
  Zadoyan, and Apkarian}}]{Apkarian01a}
\bibinfo{author}{\bibfnamefont{M.}~\bibnamefont{Karavitis}},
  \bibinfo{author}{\bibfnamefont{R.}~\bibnamefont{Zadoyan}}, \bibnamefont{and}
  \bibinfo{author}{\bibfnamefont{V.~A.} \bibnamefont{Apkarian}},
  \bibinfo{journal}{J. Chem. Phys.} \textbf{\bibinfo{volume}{{\bf 114}}},
  \bibinfo{pages}{4131} (\bibinfo{year}{{2001a}}).

\bibitem[{\citenamefont{Bihary et~al.}({2001b})\citenamefont{Bihary, Gerber,
  and Apkarian}}]{Apkarian01b}
\bibinfo{author}{\bibfnamefont{Z.}~\bibnamefont{Bihary}},
  \bibinfo{author}{\bibfnamefont{R.~B.} \bibnamefont{Gerber}},
  \bibnamefont{and} \bibinfo{author}{\bibfnamefont{V.~A.}
  \bibnamefont{Apkarian}}, \bibinfo{journal}{J. Chem. Phys.}
  \textbf{\bibinfo{volume}{{\bf 115}}}, \bibinfo{pages}{2695}
  (\bibinfo{year}{{2001b}}).

\bibitem[{\citenamefont{Bihary et~al.}({2004a})\citenamefont{Bihary, Zadoyan,
  Karavitis, and Apkarian}}]{Apkarian04a}
\bibinfo{author}{\bibfnamefont{Z.}~\bibnamefont{Bihary}},
  \bibinfo{author}{\bibfnamefont{R.}~\bibnamefont{Zadoyan}},
  \bibinfo{author}{\bibfnamefont{M.}~\bibnamefont{Karavitis}},
  \bibnamefont{and} \bibinfo{author}{\bibfnamefont{V.~A.}
  \bibnamefont{Apkarian}}, \bibinfo{journal}{J. Chem. Phys.}
  \textbf{\bibinfo{volume}{{\bf 120}}}, \bibinfo{pages}{7576}
  (\bibinfo{year}{{2004a}}).

\bibitem[{\citenamefont{Bihary et~al.}({2004b})\citenamefont{Bihary, Zadoyan,
  Karavitis, and Apkarian}}]{Apkarian04b}
\bibinfo{author}{\bibfnamefont{Z.}~\bibnamefont{Bihary}},
  \bibinfo{author}{\bibfnamefont{R.}~\bibnamefont{Zadoyan}},
  \bibinfo{author}{\bibfnamefont{M.}~\bibnamefont{Karavitis}},
  \bibnamefont{and} \bibinfo{author}{\bibfnamefont{V.~A.}
  \bibnamefont{Apkarian}}, \bibinfo{journal}{J. Chem. Phys.}
  \textbf{\bibinfo{volume}{{\bf 120}}}, \bibinfo{pages}{8144}
  (\bibinfo{year}{{2004b}}).

\bibitem[{\citenamefont{Karavitis and Apkarian}({2004c})}]{Apkarian04c}
\bibinfo{author}{\bibfnamefont{M.}~\bibnamefont{Karavitis}} \bibnamefont{and}
  \bibinfo{author}{\bibfnamefont{V.~A.} \bibnamefont{Apkarian}},
  \bibinfo{journal}{J. Chem. Phys.} \textbf{\bibinfo{volume}{{\bf 120}}},
  \bibinfo{pages}{292} (\bibinfo{year}{{2004c}}).

\bibitem[{\citenamefont{Karavitis et~al.}({2005a})\citenamefont{Karavitis,
  Kumada, Goldschleger, and Apkarian}}]{Apkarian05a}
\bibinfo{author}{\bibfnamefont{M.}~\bibnamefont{Karavitis}},
  \bibinfo{author}{\bibfnamefont{T.}~\bibnamefont{Kumada}},
  \bibinfo{author}{\bibfnamefont{I.~U.} \bibnamefont{Goldschleger}},
  \bibnamefont{and} \bibinfo{author}{\bibfnamefont{V.~A.}
  \bibnamefont{Apkarian}}, \bibinfo{journal}{PCCP}
  \textbf{\bibinfo{volume}{{\bf 7}}}, \bibinfo{pages}{791}
  (\bibinfo{year}{{2005a}}).

\bibitem[{\citenamefont{Kiviniemi et~al.}({2005b})\citenamefont{Kiviniemi,
  Aumanen, Myllyperkio, Apkarian, and Pettersson}}]{Apkarian05b}
\bibinfo{author}{\bibfnamefont{T.}~\bibnamefont{Kiviniemi}},
  \bibinfo{author}{\bibfnamefont{J.}~\bibnamefont{Aumanen}},
  \bibinfo{author}{\bibfnamefont{P.}~\bibnamefont{Myllyperkio}},
  \bibinfo{author}{\bibfnamefont{V.~A.} \bibnamefont{Apkarian}},
  \bibnamefont{and}
  \bibinfo{author}{\bibfnamefont{M.}~\bibnamefont{Pettersson}},
  \bibinfo{journal}{J. Chem. Phys.} \textbf{\bibinfo{volume}{{\bf 123}}},
  \bibinfo{pages}{064509} (\bibinfo{year}{{2005b}}).

\bibitem[{\citenamefont{Segale et~al.}({2005c})\citenamefont{Segale, Karavitis,
  Fredj, and Apkarian}}]{Apkarian05c}
\bibinfo{author}{\bibfnamefont{D.}~\bibnamefont{Segale}},
  \bibinfo{author}{\bibfnamefont{M.}~\bibnamefont{Karavitis}},
  \bibinfo{author}{\bibfnamefont{F.}~\bibnamefont{Fredj}}, \bibnamefont{and}
  \bibinfo{author}{\bibfnamefont{V.~A.} \bibnamefont{Apkarian}},
  \bibinfo{journal}{J. Chem. Phys.} \textbf{\bibinfo{volume}{{\bf 122}}},
  \bibinfo{pages}{111104} (\bibinfo{year}{{2005c}}).

\bibitem[{\citenamefont{Pechukas}({1994})}]{Pechukas94}
\bibinfo{author}{\bibfnamefont{P.}~\bibnamefont{Pechukas}},
  \bibinfo{journal}{Phys. Rev. Letts.} \textbf{\bibinfo{volume}{{\bf 73}}},
  \bibinfo{pages}{1060} (\bibinfo{year}{{1994}}).

\bibitem[{\citenamefont{Laird et~al.}({1991})\citenamefont{Laird, Budimir, and
  Skinner}}]{Laird91a}
\bibinfo{author}{\bibfnamefont{B.~B.} \bibnamefont{Laird}},
  \bibinfo{author}{\bibfnamefont{J.}~\bibnamefont{Budimir}}, \bibnamefont{and}
  \bibinfo{author}{\bibfnamefont{J.~L.} \bibnamefont{Skinner}},
  \bibinfo{journal}{J. Chem. Phys.} \textbf{\bibinfo{volume}{{\bf 94}}},
  \bibinfo{pages}{4391} (\bibinfo{year}{{1991}}).

\bibitem[{\citenamefont{Laird and Skinner}({1991})}]{Laird91b}
\bibinfo{author}{\bibfnamefont{B.~B.} \bibnamefont{Laird}} \bibnamefont{and}
  \bibinfo{author}{\bibfnamefont{J.~L.} \bibnamefont{Skinner}},
  \bibinfo{journal}{J. Chem. Phys.} \textbf{\bibinfo{volume}{{\bf 94}}},
  \bibinfo{pages}{4405} (\bibinfo{year}{{1991}}).

\bibitem[{\citenamefont{Riga and Martens}({2004})}]{Martens04}
\bibinfo{author}{\bibfnamefont{J.~M.} \bibnamefont{Riga}} \bibnamefont{and}
  \bibinfo{author}{\bibfnamefont{C.~C.} \bibnamefont{Martens}},
  \bibinfo{journal}{J. Chem. Phys.} \textbf{\bibinfo{volume}{{\bf 120}}},
  \bibinfo{pages}{6863} (\bibinfo{year}{{2004}}).

\bibitem[{\citenamefont{Riga and Martens}({2005})}]{Martens05a}
\bibinfo{author}{\bibfnamefont{J.~M.} \bibnamefont{Riga}} \bibnamefont{and}
  \bibinfo{author}{\bibfnamefont{C.~C.} \bibnamefont{Martens}},
  \bibinfo{journal}{Chem. Phys.} \textbf{\bibinfo{volume}{{\bf 322}}},
  \bibinfo{pages}{108} (\bibinfo{year}{{2005}}).

\bibitem[{\citenamefont{Riga et~al.}({2005})\citenamefont{Riga, Fredj, and
  Martens}}]{Martens05b}
\bibinfo{author}{\bibfnamefont{J.~M.} \bibnamefont{Riga}},
  \bibinfo{author}{\bibfnamefont{E.}~\bibnamefont{Fredj}}, \bibnamefont{and}
  \bibinfo{author}{\bibfnamefont{C.~C.} \bibnamefont{Martens}},
  \bibinfo{journal}{J. Chem. Phys.} \textbf{\bibinfo{volume}{{\bf 122}}},
  \bibinfo{pages}{174107} (\bibinfo{year}{{2005}}).

\bibitem[{\citenamefont{Riga et~al.}({2006})\citenamefont{Riga, Fredj, and
  Martens}}]{Martens06}
\bibinfo{author}{\bibfnamefont{J.~M.} \bibnamefont{Riga}},
  \bibinfo{author}{\bibfnamefont{E.}~\bibnamefont{Fredj}}, \bibnamefont{and}
  \bibinfo{author}{\bibfnamefont{C.~C.} \bibnamefont{Martens}},
  \bibinfo{journal}{J. Chem. Phys.} \textbf{\bibinfo{volume}{{\bf 124}}},
  \bibinfo{pages}{064506} (\bibinfo{year}{{2006}}).

\bibitem[{\citenamefont{Caldeira and Leggett}({1983a})}]{CaldeiraLeggett:QBM}
\bibinfo{author}{\bibfnamefont{A.~O.} \bibnamefont{Caldeira}} \bibnamefont{and}
  \bibinfo{author}{\bibfnamefont{A.~J.} \bibnamefont{Leggett}},
  \bibinfo{journal}{Physica} \textbf{\bibinfo{volume}{{\bf 121A}}},
  \bibinfo{pages}{587} (\bibinfo{year}{{1983a}}).

\bibitem[{\citenamefont{Caldeira and
  Leggett}({1983b})}]{CaldeiraLeggett:QTinDS}
\bibinfo{author}{\bibfnamefont{A.~O.} \bibnamefont{Caldeira}} \bibnamefont{and}
  \bibinfo{author}{\bibfnamefont{A.~J.} \bibnamefont{Leggett}},
  \bibinfo{journal}{Annals of Physics} \textbf{\bibinfo{volume}{{\bf 149}}},
  \bibinfo{pages}{374} (\bibinfo{year}{{1983b}}).

\bibitem[{\citenamefont{Caldeira and Leggett}({1985})}]{CaldeiraLeggett:Exact}
\bibinfo{author}{\bibfnamefont{A.~O.} \bibnamefont{Caldeira}} \bibnamefont{and}
  \bibinfo{author}{\bibfnamefont{A.~J.} \bibnamefont{Leggett}},
  \bibinfo{journal}{Physics Review A} \textbf{\bibinfo{volume}{{\bf 31}}},
  \bibinfo{pages}{1059} (\bibinfo{year}{{1985}}).

\bibitem[{\citenamefont{Bonella and Coker}({2005a})}]{Coker05a}
\bibinfo{author}{\bibfnamefont{S.}~\bibnamefont{Bonella}} \bibnamefont{and}
  \bibinfo{author}{\bibfnamefont{D.~F.} \bibnamefont{Coker}},
  \bibinfo{journal}{PNAS} \textbf{\bibinfo{volume}{{\bf 102}}},
  \bibinfo{pages}{6715} (\bibinfo{year}{{2005a}}).

\bibitem[{\citenamefont{Hillery et~al.}({1984})\citenamefont{Hillery,
  O'Connell, Scully, and Wigner}}]{Hillery84}
\bibinfo{author}{\bibfnamefont{M.}~\bibnamefont{Hillery}},
  \bibinfo{author}{\bibfnamefont{R.~F.} \bibnamefont{O'Connell}},
  \bibinfo{author}{\bibfnamefont{M.~O.} \bibnamefont{Scully}},
  \bibnamefont{and} \bibinfo{author}{\bibfnamefont{E.~P.}
  \bibnamefont{Wigner}}, \bibinfo{journal}{Phys. Rep.}
  \textbf{\bibinfo{volume}{{\bf 106}}}, \bibinfo{pages}{121}
  (\bibinfo{year}{{1984}}).

\bibitem[{\citenamefont{Mukamel}({1982})}]{Mukamel82}
\bibinfo{author}{\bibfnamefont{S.}~\bibnamefont{Mukamel}}, \bibinfo{journal}{J.
  Chem. Phys.} \textbf{\bibinfo{volume}{{\bf 77}}}, \bibinfo{pages}{173}
  (\bibinfo{year}{{1982}}).

\bibitem[{\citenamefont{Shemetulskis and Loring}({1992})}]{Shemetulskis92}
\bibinfo{author}{\bibfnamefont{N.~E.} \bibnamefont{Shemetulskis}}
  \bibnamefont{and} \bibinfo{author}{\bibfnamefont{R.~F.}
  \bibnamefont{Loring}}, \bibinfo{journal}{J. Chem. Phys.}
  \textbf{\bibinfo{volume}{{\bf 97}}}, \bibinfo{pages}{1217}
  (\bibinfo{year}{{1992}}).

\bibitem[{\citenamefont{Egorov et~al.}({1998})\citenamefont{Egorov, Rabani, and
  Berne}}]{Egorov98}
\bibinfo{author}{\bibfnamefont{S.~A.} \bibnamefont{Egorov}},
  \bibinfo{author}{\bibfnamefont{E.}~\bibnamefont{Rabani}}, \bibnamefont{and}
  \bibinfo{author}{\bibfnamefont{B.~J.} \bibnamefont{Berne}},
  \bibinfo{journal}{J. Chem. Phys.} \textbf{\bibinfo{volume}{{\bf 108}}},
  \bibinfo{pages}{1407} (\bibinfo{year}{{1998}}).

\bibitem[{\citenamefont{Poulsen and Nyman}({2004})}]{Poulsen04}
\bibinfo{author}{\bibfnamefont{J.~A.} \bibnamefont{Poulsen}} \bibnamefont{and}
  \bibinfo{author}{\bibfnamefont{G.}~\bibnamefont{Nyman}}, \bibinfo{journal}{J.
  Phys. Chem.} \textbf{\bibinfo{volume}{{\bf 108}}}, \bibinfo{pages}{8743}
  (\bibinfo{year}{{2004}}).

\bibitem[{\citenamefont{Poulsen et~al.}({2005})\citenamefont{Poulsen, Nyman,
  and Rossky}}]{Poulsen05}
\bibinfo{author}{\bibfnamefont{J.~A.} \bibnamefont{Poulsen}},
  \bibinfo{author}{\bibfnamefont{G.}~\bibnamefont{Nyman}}, \bibnamefont{and}
  \bibinfo{author}{\bibfnamefont{P.~J.} \bibnamefont{Rossky}},
  \bibinfo{journal}{PNAS} \textbf{\bibinfo{volume}{{\bf 102}}},
  \bibinfo{pages}{6709} (\bibinfo{year}{{2005}}).

\bibitem[{\citenamefont{Cao and Voth}({1993})}]{Cao93}
\bibinfo{author}{\bibfnamefont{J.}~\bibnamefont{Cao}} \bibnamefont{and}
  \bibinfo{author}{\bibfnamefont{G.~A.} \bibnamefont{Voth}},
  \bibinfo{journal}{J. Chem. Phys.} \textbf{\bibinfo{volume}{{\bf 99}}},
  \bibinfo{pages}{10070} (\bibinfo{year}{{1993}}).

\bibitem[{\citenamefont{Cao and Voth}({1994a})}]{Cao94a}
\bibinfo{author}{\bibfnamefont{J.}~\bibnamefont{Cao}} \bibnamefont{and}
  \bibinfo{author}{\bibfnamefont{G.~A.} \bibnamefont{Voth}},
  \bibinfo{journal}{J. Chem. Phys.} \textbf{\bibinfo{volume}{{\bf 100}}},
  \bibinfo{pages}{5093} (\bibinfo{year}{{1994a}}).

\bibitem[{\citenamefont{Cao and Voth}({1994b})}]{Cao94b}
\bibinfo{author}{\bibfnamefont{J.}~\bibnamefont{Cao}} \bibnamefont{and}
  \bibinfo{author}{\bibfnamefont{G.~A.} \bibnamefont{Voth}},
  \bibinfo{journal}{J. Chem. Phys.} \textbf{\bibinfo{volume}{{\bf 100}}},
  \bibinfo{pages}{5106} (\bibinfo{year}{{1994b}}).

\bibitem[{\citenamefont{Cao and Voth}({1994c})}]{Cao94c}
\bibinfo{author}{\bibfnamefont{J.}~\bibnamefont{Cao}} \bibnamefont{and}
  \bibinfo{author}{\bibfnamefont{G.~A.} \bibnamefont{Voth}},
  \bibinfo{journal}{J. Chem. Phys.} \textbf{\bibinfo{volume}{{\bf 101}}},
  \bibinfo{pages}{6157} (\bibinfo{year}{{1994c}}).

\bibitem[{\citenamefont{Cao and Voth}({1994d})}]{Cao94d}
\bibinfo{author}{\bibfnamefont{J.}~\bibnamefont{Cao}} \bibnamefont{and}
  \bibinfo{author}{\bibfnamefont{G.~A.} \bibnamefont{Voth}},
  \bibinfo{journal}{J. Chem. Phys.} \textbf{\bibinfo{volume}{{\bf 101}}},
  \bibinfo{pages}{6168} (\bibinfo{year}{{1994d}}).

\bibitem[{\citenamefont{Cao and Voth}({1994e})}]{Cao94e}
\bibinfo{author}{\bibfnamefont{J.}~\bibnamefont{Cao}} \bibnamefont{and}
  \bibinfo{author}{\bibfnamefont{G.~A.} \bibnamefont{Voth}},
  \bibinfo{journal}{J. Chem. Phys.} \textbf{\bibinfo{volume}{{\bf 101}}},
  \bibinfo{pages}{6184} (\bibinfo{year}{{1994e}}).

\bibitem[{\citenamefont{Cao and Berne}({1990})}]{Cao90}
\bibinfo{author}{\bibfnamefont{J.}~\bibnamefont{Cao}} \bibnamefont{and}
  \bibinfo{author}{\bibfnamefont{B.}~\bibnamefont{Berne}}, \bibinfo{journal}{J.
  Chem. Phys.} \textbf{\bibinfo{volume}{{\bf 92}}}, \bibinfo{pages}{7531}
  (\bibinfo{year}{{1990}}).

\bibitem[{\citenamefont{Feynman and Kleinert}({1986})}]{Kleinert86}
\bibinfo{author}{\bibfnamefont{R.~P.} \bibnamefont{Feynman}} \bibnamefont{and}
  \bibinfo{author}{\bibfnamefont{H.}~\bibnamefont{Kleinert}},
  \bibinfo{journal}{Phys. Rev. A} \textbf{\bibinfo{volume}{{\bf 34}}},
  \bibinfo{pages}{5080} (\bibinfo{year}{{1986}}).

\bibitem[{\citenamefont{Janke and Kleinert}({1987})}]{Kleinert87}
\bibinfo{author}{\bibfnamefont{W.}~\bibnamefont{Janke}} \bibnamefont{and}
  \bibinfo{author}{\bibfnamefont{H.}~\bibnamefont{Kleinert}},
  \bibinfo{journal}{Chem. Phys. Lett.} \textbf{\bibinfo{volume}{{\bf 137}}},
  \bibinfo{pages}{162} (\bibinfo{year}{{1987}}).

\bibitem[{\citenamefont{Roepstorff}({1994})}]{Roepstorff94}
\bibinfo{author}{\bibfnamefont{G.}~\bibnamefont{Roepstorff}},
  \emph{\bibinfo{title}{Path Integral Approach to Quantum Physics An
  Introduction}} (\bibinfo{publisher}{Springer-Verlag}, \bibinfo{address}{New
  York}, \bibinfo{year}{{1994}}).

\bibitem[{\citenamefont{Kleinert}({2004})}]{Kleinert04}
\bibinfo{author}{\bibfnamefont{H.}~\bibnamefont{Kleinert}},
  \emph{\bibinfo{title}{Path Integrals in Quantum Mechanics,Statistics, Polymer
  Physics, and Financial Markets}} (\bibinfo{publisher}{World Scientific},
  \bibinfo{address}{Singapore}, \bibinfo{year}{{2004}}).

\bibitem[{\citenamefont{Feynman}({1998})}]{Feynman98}
\bibinfo{author}{\bibfnamefont{R.~P.} \bibnamefont{Feynman}},
  \emph{\bibinfo{title}{Statistical Mechanics,a set of lectures}}
  (\bibinfo{publisher}{Westview Press}, \bibinfo{address}{Colorado},
  \bibinfo{year}{{1998}}).

\bibitem[{\citenamefont{Feynman and Hibbs}({1965})}]{Feynman-Hibbs65}
\bibinfo{author}{\bibfnamefont{R.~P.} \bibnamefont{Feynman}} \bibnamefont{and}
  \bibinfo{author}{\bibfnamefont{A.~R.} \bibnamefont{Hibbs}},
  \emph{\bibinfo{title}{Quantum mechanics and path integrals}}
  (\bibinfo{publisher}{McGraw-Hill}, \bibinfo{address}{New York},
  \bibinfo{year}{{1965}}).

\bibitem[{\citenamefont{Shi and Geva}({2003a})}]{ShiGeva03a}
\bibinfo{author}{\bibfnamefont{Q.}~\bibnamefont{Shi}} \bibnamefont{and}
  \bibinfo{author}{\bibfnamefont{E.}~\bibnamefont{Geva}}, \bibinfo{journal}{J.
  Phys. Chem.} \textbf{\bibinfo{volume}{{\bf 107}}}, \bibinfo{pages}{9059}
  (\bibinfo{year}{{2003a}}).

\bibitem[{\citenamefont{Klemm and Storer}({1973})}]{Storer73}
\bibinfo{author}{\bibfnamefont{A.}~\bibnamefont{Klemm}} \bibnamefont{and}
  \bibinfo{author}{\bibfnamefont{R.~G.} \bibnamefont{Storer}},
  \bibinfo{journal}{Aus. J. Phys.} \textbf{\bibinfo{volume}{{\bf 26}}},
  \bibinfo{pages}{43} (\bibinfo{year}{{1973}}).

\bibitem[{\citenamefont{Thirumalai et~al.}({1983})\citenamefont{Thirumalai,
  Bruskin, and Berne}}]{Thirumalai83}
\bibinfo{author}{\bibfnamefont{D.}~\bibnamefont{Thirumalai}},
  \bibinfo{author}{\bibfnamefont{E.~J.} \bibnamefont{Bruskin}},
  \bibnamefont{and} \bibinfo{author}{\bibfnamefont{B.~J.} \bibnamefont{Berne}},
  \bibinfo{journal}{J. Chem. Phys.} \textbf{\bibinfo{volume}{{\bf 79}}},
  \bibinfo{pages}{5063} (\bibinfo{year}{{1983}}).

\bibitem[{\citenamefont{Karsch et~al.}({1984})\citenamefont{Karsch, Rabinovich,
  shore, and Veneziano}}]{Karsch84}
\bibinfo{author}{\bibfnamefont{F.}~\bibnamefont{Karsch}},
  \bibinfo{author}{\bibfnamefont{E.}~\bibnamefont{Rabinovich}},
  \bibinfo{author}{\bibfnamefont{G.}~\bibnamefont{shore}}, \bibnamefont{and}
  \bibinfo{author}{\bibfnamefont{G.}~\bibnamefont{Veneziano}},
  \bibinfo{journal}{Nuc. Phys. B} \textbf{\bibinfo{volume}{{\bf 242}}},
  \bibinfo{pages}{503} (\bibinfo{year}{{1984}}).

\bibitem[{\citenamefont{Meier and Beswick}({2004})}]{Beswick04a}
\bibinfo{author}{\bibfnamefont{C.}~\bibnamefont{Meier}} \bibnamefont{and}
  \bibinfo{author}{\bibfnamefont{J.~A.} \bibnamefont{Beswick}},
  \bibinfo{journal}{J. Chem. Phys.} \textbf{\bibinfo{volume}{{\bf 121}}},
  \bibinfo{pages}{4550} (\bibinfo{year}{{2004}}).

\bibitem[{\citenamefont{Lohmuller et~al.}({2004})\citenamefont{Lohmuller,
  Engel, Meier, and Beswick}}]{Beswick04b}
\bibinfo{author}{\bibfnamefont{T.}~\bibnamefont{Lohmuller}},
  \bibinfo{author}{\bibfnamefont{V.}~\bibnamefont{Engel}},
  \bibinfo{author}{\bibfnamefont{C.}~\bibnamefont{Meier}}, \bibnamefont{and}
  \bibinfo{author}{\bibfnamefont{J.~A.} \bibnamefont{Beswick}},
  \bibinfo{journal}{J. Chem. Phys.} \textbf{\bibinfo{volume}{{\bf 120}}},
  \bibinfo{pages}{10442} (\bibinfo{year}{{2004}}).

\bibitem[{\citenamefont{Herzberg}({1950})}]{Herzberg50}
\bibinfo{author}{\bibfnamefont{G.}~\bibnamefont{Herzberg}},
  \emph{\bibinfo{title}{Molecular Spectra and Molecular Structure I. Spectra of
  Diatomic Molecules}} (\bibinfo{publisher}{D. Van Noshtrand Company},
  \bibinfo{address}{New York}, \bibinfo{year}{{1950}}).

\bibitem[{\citenamefont{Allen and Tildesley}({1987})}]{Allen87}
\bibinfo{author}{\bibfnamefont{M.~P.} \bibnamefont{Allen}} \bibnamefont{and}
  \bibinfo{author}{\bibfnamefont{D.~J.} \bibnamefont{Tildesley}},
  \emph{\bibinfo{title}{Computer Simulation of Liquids}}
  (\bibinfo{publisher}{Oxford Science Publications}, \bibinfo{address}{New
  York}, \bibinfo{year}{{1987}}).

\bibitem[{\citenamefont{Shugard et~al.}({1978})\citenamefont{Shugard, Tully,
  and Nitzan}}]{Nitzan78}
\bibinfo{author}{\bibfnamefont{M.}~\bibnamefont{Shugard}},
  \bibinfo{author}{\bibfnamefont{J.}~\bibnamefont{Tully}}, \bibnamefont{and}
  \bibinfo{author}{\bibfnamefont{A.}~\bibnamefont{Nitzan}},
  \bibinfo{journal}{J. Chem. Phys.} \textbf{\bibinfo{volume}{{\bf 69}}},
  \bibinfo{pages}{336} (\bibinfo{year}{{1978}}).

\bibitem[{\citenamefont{Nitzan et~al.}({1978})\citenamefont{Nitzan, Shugard,
  and Tully}}]{Nitzan78a}
\bibinfo{author}{\bibfnamefont{A.}~\bibnamefont{Nitzan}},
  \bibinfo{author}{\bibfnamefont{M.}~\bibnamefont{Shugard}}, \bibnamefont{and}
  \bibinfo{author}{\bibfnamefont{J.}~\bibnamefont{Tully}}, \bibinfo{journal}{J.
  Chem. Phys.} \textbf{\bibinfo{volume}{{\bf 69}}}, \bibinfo{pages}{2525}
  (\bibinfo{year}{{1978}}).

\bibitem[{\citenamefont{Harris}({1977})}]{Harris77}
\bibinfo{author}{\bibfnamefont{C.}~\bibnamefont{Harris}}, \bibinfo{journal}{J.
  Chem. Phys.} \textbf{\bibinfo{volume}{{\bf 67}}}, \bibinfo{pages}{5607}
  (\bibinfo{year}{{1977}}).

\bibitem[{\citenamefont{Shelby et~al.}({1979})\citenamefont{Shelby, Harris, and
  Cornelius}}]{Harris79}
\bibinfo{author}{\bibfnamefont{R.}~\bibnamefont{Shelby}},
  \bibinfo{author}{\bibfnamefont{C.}~\bibnamefont{Harris}}, \bibnamefont{and}
  \bibinfo{author}{\bibfnamefont{P.}~\bibnamefont{Cornelius}},
  \bibinfo{journal}{J. Chem. Phys.} \textbf{\bibinfo{volume}{{\bf 70}}},
  \bibinfo{pages}{34} (\bibinfo{year}{{1979}}).

\bibitem[{\citenamefont{Marks et~al.}({1980})\citenamefont{Marks, Corneliu, and
  Harris}}]{Harris80}
\bibinfo{author}{\bibfnamefont{S.}~\bibnamefont{Marks}},
  \bibinfo{author}{\bibfnamefont{P.}~\bibnamefont{Corneliu}}, \bibnamefont{and}
  \bibinfo{author}{\bibfnamefont{C.}~\bibnamefont{Harris}},
  \bibinfo{journal}{J. Chem. Phys.} \textbf{\bibinfo{volume}{{\bf 73}}},
  \bibinfo{pages}{3069} (\bibinfo{year}{{1980}}).

\bibitem[{\citenamefont{deBree and Wiersma}({1979})}]{Wiersma79}
\bibinfo{author}{\bibfnamefont{P.}~\bibnamefont{deBree}} \bibnamefont{and}
  \bibinfo{author}{\bibfnamefont{D.}~\bibnamefont{Wiersma}},
  \bibinfo{journal}{J. Chem. Phys.} \textbf{\bibinfo{volume}{{\bf 70}}},
  \bibinfo{pages}{790} (\bibinfo{year}{{1979}}).

\bibitem[{\citenamefont{Ma and Coker}({2008})}]{Ma08}
\bibinfo{author}{\bibfnamefont{Z.}~\bibnamefont{Ma}} \bibnamefont{and}
  \bibinfo{author}{\bibfnamefont{D.~F.} \bibnamefont{Coker}}
  (\bibinfo{year}{{2008}}).

\bibitem[{\citenamefont{Coker}({2008})}]{Coker08}
\bibinfo{author}{\bibfnamefont{D.~F.} \bibnamefont{Coker}}
  (\bibinfo{year}{{2008}}).

\end{thebibliography}

\end{document}